\documentclass[prb,twocolumn,eqsecnum,superscriptaddress]{revtex4}
\usepackage{bm,times,amsmath,amssymb,amsfonts}
\usepackage[usenames,dvipsnames]{pstricks}
\usepackage{graphicx}

\begin{document}

\title{Dynamic critical behavior of model $\bm{A}$ in films: Zero-mode
  boundary conditions and
  expansion near four dimensions}
\author{H. W. Diehl}
\affiliation{Fachbereich Physik, Universit\"at Duisburg-Essen,
 Lotharstra{\ss}e 1,
47048 Duisburg, Germany}
\affiliation{Kavli Institute for Theoretical Physics, University of
  California--Santa Barbara, Santa Barbara, California 93106-4030, USA}
\author{H. Chamati}
\affiliation{Fachbereich Physik, Universit\"at Duisburg-Essen,
 Lotharstra{\ss}e 1,
47048 Duisburg, Germany}
\affiliation{Institute of Solid State Physics, Bulgarian Academy of Sciences,
72 Tzarigradsko Chauss\'{e}e, 1784 Sofia,
Bulgaria}
\date{\today}
\begin{abstract}
  The critical dynamics of relaxational stochastic models with
  nonconserved $n$-component order parameter $\bm{\phi}$ and no
  coupling to other slow variables (``model $A$'') is investigated in
  film geometries for the cases of periodic and free boundary
  conditions. The Hamiltonian $\mathcal{H}$ governing the stationary
  equilibrium distribution is taken to be $O(n)$ symmetric and to
  involve, in the case of free boundary conditions, the boundary terms
  $\int_{\mathfrak{B}_j}\mathring{c}_j\,\phi^2/2$ associated with the
  two confining surface planes $\mathfrak{B}_j$, $j=1,2$, at $z=0$ and
  $z=L$. Both enhancement variables $\mathring{c}_j$ are presumed to
  be subcritical or critical, so that no long-range surface order can
  occur above the bulk critical temperature $T_{c,\infty}$. A
  field-theoretic renormalization-group study of the dynamic critical
  behavior at $d=4-\epsilon$ bulk dimensions is presented, with
  special attention paid to the cases where the classical theories
  involve zero modes at $T_{c,\infty}$. This applies when either both
  $\mathring{c}_j$ take the critical value $\mathring{c}_{\text{sp}}$
  associated with the special surface transition or else periodic
  boundary conditions are imposed. Owing to the zero modes, the
  $\epsilon$~expansion becomes ill-defined at $T_{c,\infty}$.
  Analogously to the static case, the field theory can be reorganized
  to obtain a well-defined small-$\epsilon$ expansion involving
  half-integer powers of $\epsilon$, modulated by powers of
  $\ln\epsilon$. This is achieved through the construction of an
  effective ($d-1$)-dimensional action for the zero-mode component of
  the order parameter by integrating out its orthogonal component via
  renormalization-group improved perturbation theory. Explicit results
  for the scaling functions of temperature-dependent finite-size
  susceptibilities at temperatures $T\ge T_{c,\infty}$ and of layer
  and surface susceptibilities at the bulk critical point are given to
  orders $\epsilon$ and $\epsilon^{3/2}$, respectively. They show that
  $L$~dependent shifts of the multicritical special point occur along
  the temperature and enhancement axes. For the case of periodic
  boundary conditions, the consistency of the expansions to
  $O(\epsilon^{3/2})$ with exact large-$n$ results is shown. We also
  discuss briefly the effects of weak anisotropy, relating theories
  whose Hamiltonian involves a generalized square gradient term $B^{kl}
  \partial_k\bm{\phi}\cdot\partial_l\bm{\phi}$ to those with a conventional $(\nabla\bm{\phi})^2$
  term.
\end{abstract}
\preprint{NSF-KITP-08-127}
\maketitle

\section{Introduction}
The renormalization group (RG) approach has played an important role
in the modern theory of critical phenomena.
\cite{RGrev,WK74,Weg76,Fis98} For one thing, it provides an
appropriate mathematical framework for the formulation of the theory.
Second, it has led to the development of powerful calculational tools
for quantitatively accurate investigations. Its most impressive and
numerous successes have been achieved in the study of static bulk
critical phenomena. However, appropriate extensions for studies of
dynamic bulk critical phenomena, \cite{HH77,FM06} boundary critical
phenomena,\cite{Die86a,Die97} and finite-size effects
\cite{remFSE,Sym81,Bre82,BZJ85,RGJ85,Bar83,Pri90,BDT00,ZJ02,Gol87a,NZ-J87,Die87,Doh89,Doh93}
were developed, which have proven their utility and power. Anyone of
the features, ``dynamics'', ``boundaries'', and ``finite size,''
involves fundamental new issues and adds to the technical complexity
of analytic RG studies. It is therefore not surprising that work on
problems involving combinations of several of these features has
remained rather scarce.

In this paper we shall be concerned with the dynamics of systems in
slabs $\mathbb{R}^{d-1}\times[0,L]$ of finite thickness $L$ near their
bulk ($L=\infty$) critical point. Both the cases of periodic and free
boundary conditions will be considered. Hence we shall have to deal
with all three of the above-mentioned features.

Despite the availability of some experimental results, there exist only few
previous studies of dynamic critical behavior of systems in film
geometry (see Ref.~\onlinecite{BHLD07} and its references).  In two
earlier papers, Calvo and Ferrell \cite{CF85,Cal85}  investigated the dynamics of binary
liquid mixtures confined between two parallel plates using the
mode-mode coupling approach. Subsequently, the
dynamics of bounded one-component fluids near the liquid-gas critical
point (model $H$ in the terminology of Ref.~\onlinecite{HH77}) and of
confined liquid $^4$He at the superfluid transition (model $E$) were
analyzed by this method.\cite{Bha96} There exist also a number of
papers dealing with finite-size effects on dynamic critical behavior
and dynamic surface critical behavior.
\cite{DD83b,Die87,Gol87a,Gol87b,NZ-J87,DD85,FD89,XG89a,XG89b,FD91,BF91,DR93,DJ92,Doh93,Die94b,WD95,RC95,RD96,KKD01,DKK02,MdVAMB07}
In other works, finite-size systems with long-range interactions or
quenched disorder were investigated.\cite{chamati2001a,chamati2002a,Gam08}

Recently, Gambassi and Dietrich \cite{GD06b} presented a
fairly detailed study of the familiar model $A$
(Refs.~\onlinecite{HH77} and \onlinecite{FM06}) in film geometry within the framework of
the classical (zero-loop, van Hove) approximation, augmented by
RG-improved perturbation theory. They focused on the situation where
the surface interactions on both confining surfaces are subcritically
enhanced. This corresponds to the case in which ordinary surface
transitions \cite{Die86a,Bin83} occur in the semi-infinite systems
bounded by either one of the two surface planes.  Knowing that
Dirichlet boundary conditions apply under these conditions at both
boundary planes on sufficiently large length scales, they restricted
their analysis by choosing such boundary conditions from the outset.

In the present paper we shall also be concerned with model $A$ in film
geometry. Our analysis complements and goes beyond that of
Ref.~\onlinecite{GD06b} in several ways. First, we shall not limit
ourselves to the classical approximation but present a RG analysis in
$d=4-\epsilon$ bulk dimensions, going to one-loop order in our
explicit calculations (and partly beyond to determine contributions of
order $\epsilon^{3/2}$). Second, we shall give up the restriction to
Dirichlet boundary conditions on both confining plates.  Aside from
periodic boundary conditions, we shall consider, in the general part
of our analysis, the generic case of symmetry preserving Robin
boundary conditions corresponding to distinct enhancements of the
surface interactions on the two boundary planes. This includes the
case of special-special (sp-sp) boundary conditions for which the
surface interactions on both boundary planes are critically enhanced.
\cite{KD92a,DGS06,GD08} 

Periodic and sp-sp boundary conditions share the feature that Landau
theory involves a zero mode at bulk criticality. It has recently
become clear that this causes a breakdown of the $\epsilon$ expansion
at $T_{c,\infty}$.  \cite{DGS06,GD08} The small-$\epsilon$ expansions
of the associated universal amplitudes of the critical Casimir forces
were found to involve, besides integer powers of $\epsilon$, also
fractional powers $\epsilon^{k/2}$ with $k\ge 3$ (modulo powers of
$\ln\epsilon$). This breakdown of the $\epsilon$ expansion is similar
to the one reported in Ref.~\onlinecite{Sac97} for the $3-d$ expansion
of bosonic quantum systems.

The primary aim of this paper is to show that a similar breakdown of
the $\epsilon$ expansion is encountered in the study of dynamic
critical behavior of model $A$ in film geometry for periodic and sp-sp
boundary conditions. We shall demonstrate this explicitly by
determining the contributions of order $\epsilon^{3/2}$ of the dynamic
finite-size susceptibility $\chi_L$ for both boundary conditions, the
layer susceptibility $\chi^{(\text{per})}_{zz}(L)$ for periodic
boundary conditions, and their surface analogs
$\chi^{(\text{sp-sp})}_{11}(L)$ and $\chi^{(\text{sp-sp})}_{1L}(L)$
for sp-sp boundary conditions at $T_{c,\infty}$.

The remainder of this paper is organized as follows. In Sec.~II,
we first define an appropriate extension of model $A$ to the
film geometry. We then recall the Lagrangian formulation of the
corresponding Langevin equation \cite{Jan76,dDom76,Jan92,DD83b} along
with some necessary background such as the fluctuation-dissipation
theorem and the boundary conditions of the order-parameter field
$\bm{\phi}$ and the associated response field $\tilde{\bm{\phi}}$.
\cite{DJ92,Die94b} In Sec.~\ref{sec:PRRen}, we set up perturbation
theory, explain the renormalization of the theory for $d=4-\epsilon$,
give the RG equations of the multi-point correlation and response
functions, and describe their solutions.  Section~\ref{sec:RGipt}
begins with a discussion of the basis of RG-improved perturbation
theory. Next, we show that the $\epsilon$ expansion breaks down at
$T_{c,\infty}$ for periodic and sp-sp boundary conditions and
elucidate the origin of the problem. To obtain well-defined
small-$\epsilon$ expansions, we then construct an effective
($d-1$)-dimensional dynamic field theory for the zero-mode components
of $\bm{\phi}$ and $\tilde{\bm{\phi}}$. In Sec.~\ref{sec:scffss} we
present one-loop results for various scaling functions for $T\ge
T_{c,\infty}$.  Section~\ref{sec:concl} contains a brief summary and
concluding remarks. In addition, we briefly embark on the issue of
universality violations due to weak anisotropy and other sources
brought up recently.\cite{CD02b,Doh08} Finally, there are three
appendixes in which technical details are described.

\section{The model and background}\label{sec:smodel}
\subsection{Definition of model $A$ in film geometry}

We begin by defining an appropriate extension of model $A$ for the film
geometry.  To this end, we consider a film occupying the region
$\mathfrak{V}=\mathbb{R}^{d-1}\times [0,L]$ of $d$-dimensional space
$\mathbb{R}^d$. We wish to study the critical dynamics of such films
involving an $n$-component order-parameter field
$\bm{\phi}(\bm{x},t)=(\phi_\alpha(\bm{x},t),\alpha=1,\dotsc,n)$. We
write position vectors as $\bm{x}=(\bm{y},z)$, where
$\bm{y}\in\mathbb{R}^{d-1}$ and $z\in[0,L]$ are the coordinates
alongside and across the film, respectively. We choose periodic
boundary conditions along the $d-1$ principal $y$~directions.
Depending on whether we are concerned with periodic or free boundary
conditions in the $z$-direction, the slab $\mathfrak{V}$ has no
boundary, $\partial\mathfrak{V}=\emptyset$, or consists of the two
($d-1$)-dimensional confining hyperplanes $\mathfrak{B}_1$ at $z=0$
and $\mathfrak{B}_2$ at $z=L$. In the latter case, we orient the
boundary such that the normal $\bm{n}$ on
$\partial\mathfrak{V}=\mathfrak{B}\equiv\mathfrak{B}_1\cup\mathfrak{B}_2$
points in the interior of $\mathfrak{V}$.

We are interested in the dynamics of systems that relax to a 
stationary equilibrium state described by the Hamiltonian (in units of
$k_BT$)
\begin{eqnarray}\label{eq:Ham}
\mathcal{H}[\bm{\phi}]&=&\int_{\mathfrak{V}}d^dx\,
\bigg[\frac{1}{2}\left(\nabla\bm{\phi}\right)^2+\frac{\mathring{\tau}}2\phi^2
+\frac{\mathring{u}}{4!}\phi^4\bigg]\nonumber\\ &&\strut
+\delta_{\wp,f} \sum_{j=1}^2 \int_{\mathfrak{B}_j}d^{d-1}y\,
\frac{\mathring{c}_j}{2}\,\phi^2 \;.
\end{eqnarray}
Here the contributions localized on the boundary planes
$\mathfrak{B}_j$, given in the second line of Eq.~\eqref{eq:Ham}, are
only present for free boundary conditions ($\wp=f$) but
absent for periodic boundary conditions ($\wp=\mathrm{per}$). The
absence of boundary terms linear in $\bm{\phi}$ reflects our
assumption that the boundaries do not break the
$\bm{\phi}\to-\bm{\phi}$ of the Hamiltonian. That no quadratic
anisotropies have been taken into account in the boundary terms meets
our stronger requirement that the boundaries do not break the presumed
$O(n)$ symmetry. Surface spin anisotropies, which would require
separate enhancement variables $\mathring{c}_{j,\alpha}$ for the boundary terms
$\propto \phi_\alpha^2$ of different components $\alpha$,
\cite{DE82,DE84} will not be considered here. 

For the sake of simplicity, we shall furthermore assume that the values of
both surface variables $\mathring{c}_j$ are such that no long-range surface
order can occur at $\mathfrak{B}_j$ above the bulk critical temperature
$T_{c,\infty}$. Recall that in a semi-infinite system bounded by
$\mathfrak{B}_j$, a transition to a bulk-disordered, surface-ordered
phase takes place at a temperature $T_{c,s}>T_{c,\infty}$ when $\mathring{c}_j$
drops below the threshold value $\mathring{c}_{\mathrm{sp}}$ associated with
the so-called special transition (provided the dimension $d$ is
sufficiently large that the $d-1$~dimensional surface can support
long-range order). Thus, our assumption translates into the conditions
\begin{equation}
  \delta\mathring{c}_j\equiv\mathring{c}_j-\mathring{c}_{\mathrm{sp}}\ge 0\;,\quad j=1,2\;;
\end{equation}
their physical meaning is that the surface pair interactions are
subcritically ($\delta\mathring{c}_j>0$) or critically ($\delta\mathring{c}_j=0$) but not
supercritically ($\delta\mathring{c}_j<0$) enhanced. 

We are now ready to define an appropriate extension of model $A$ to the
film geometry considered that is compatible with our assumptions.
Straightforward considerations analogous to those made in
Refs.~\onlinecite{DJ92} and \onlinecite{Die94b} for
semi-infinite systems lead us to consider the Langevin equations
\begin{subequations}\label{eq:langevin}
\begin{equation}
\dot{\phi}_\alpha(\bm{x},t)=-\mathring{\lambda}\,\frac{\delta
\mathcal{H}}{\delta\phi_\alpha}(\bm{x},t)+\zeta_\alpha(\bm{x},t)\;,
\end{equation}
which are meant in the sense of Ito.\cite{vKa92} Here
$\mathring{\lambda}$ is the bare Onsager coefficient and $\bm{\zeta}$ is a
Gaussian random force of mean zero,
\begin{equation}
  \label{eq:avzeta}
  \langle\zeta_\alpha(\bm{x},t)\rangle=0\;,
\end{equation} and variance
\begin{equation}\label{gauss}
\langle\zeta_\alpha(\bm{x},t)\,\zeta_{\alpha'}(\bm{x}',t')\rangle=
2\mathring{\lambda}\,\delta_{\alpha\alpha'}\,\delta(\bm{x}-\bm{x}')\,\delta(t-t')\;.
\end{equation}
\end{subequations}

\subsection{Lagrangian formulation of the theory}
\label{sec:lagform}

In our subsequent analysis of this model it will be convenient to use
its equivalent Lagrangian formulation.
\cite{Jan76,dDom76,Jan92,DD83b,DJ92,Die94b} This
involves the action
\begin{eqnarray}\label{eq:J}
\mathcal{J}[\tilde{\bm{\phi}},\bm{\phi}]&=&
\int_{t_i}^{t_f}\!dt\,\bigg[\int_{\mathfrak{V}}
\bigg\{\tilde{\bm{\phi}}\cdot\bigg[\mathring{\lambda}\,
\bigg(\overleftarrow{\nabla}\cdot\overrightarrow{\nabla}+\mathring{\tau} 
+\frac{\mathring{u}}{3!}\,\phi^2\bigg)\bm{\phi} 
\nonumber\\ &&\strut 
+\dot{\bm{\phi}}-\mathring{\lambda}\,\tilde{\bm{\phi}}\bigg]
\bigg\}+\delta_{\wp,f} \mathring{\lambda}\sum_{j=1}^2 \int_{\mathfrak{B}_j}\,
\mathring{c}_j\,\tilde{\bm{\phi}}\cdot\bm{\phi}\bigg],\phantom{(2.4)}
\end{eqnarray}
where $\tilde{\bm{\phi}}$ is an auxiliary field, the so-called response
field. The gradient operators $\overleftarrow{\nabla}$ and
$\overrightarrow{\nabla}$ act as indicated to the left and right,
respectively.\cite{DJ92,Die94b,rem:nabla} Note that we 
have dropped a contribution  $\propto\theta(t=0)$ [where $\theta(t)$
is the Heaviside function] produced by the
Jacobian $\det(\delta\zeta_\alpha/\delta\phi_\beta)$, choosing a
prepoint discretization in time.

We fix the initial condition for the solutions to
Eq.~\eqref{eq:langevin} in the infinite past, taking the limits
$t_i\to-\infty$ and $t_f\to\infty$, and suppress these integration
limits henceforth. Multipoint correlation functions of the fields
$\phi_\alpha(\bm{x},t)$ and
$\tilde{\phi}_{\tilde{\alpha}}(\tilde{\bm{x}},\tilde{t})$ can then be
calculated with the functional weight
$\exp(-\mathcal{J}[\tilde{\bm{\phi}},\bm{\phi}])\,\mathcal{D}[\tilde{\bm{\phi}},\bm{\phi}]$,
where the measure $\mathcal{D}[\tilde{\bm{\phi}}, \bm{\phi}]$ is
proportional to $\prod_{\bm{x},\alpha,t} d(\tilde{\phi}_\alpha/2\pi
i)\,d\phi_\alpha$ and normalized such that
\begin{equation}
  \label{eq:Jnorm}
 \int\mathcal{D}[\tilde{\bm{\phi}},\bm{\phi}]\,
 e^{-\mathcal{J}[\tilde{\bm{\phi}},\bm{\phi}]}=1\;. 
\end{equation}

The field $\tilde{\phi}_\alpha(\bm{x},t)$ describes how averages
$\langle\mathcal{O}[\bm{\phi}]\rangle $ of observables
$\mathcal{O}[\bm{\phi}]$ obtained from the solutions to the Langevin
equation~\eqref{eq:langevin} upon averaging over noise histories
respond to perturbations that change its right-hand side by a function
$\tilde{J}_\alpha(\bm{x},t)$. In the case of model $A$, the addition of
the time-dependent magnetic-field terms
\begin{eqnarray}
  \label{eq:Hh}
  \mathcal{H}_{\text{fields}} &=&-\int_{\mathfrak{V}}d^dx\,
  \mathring{\bm{h}}(\bm{x},t)\cdot\bm{\phi}(\bm{x},t)\nonumber\\ &&\strut
-\delta_{\wp,f}\int_{\mathfrak{B}}d^{d-1}y\,
\mathring{\bm{h}}^{\mathfrak{B}}(\bm{x},t)\cdot\bm{\phi}(\bm{x},t) 
\end{eqnarray}
to Hamiltonian~\eqref{eq:Ham} would yield such a perturbation with
$\tilde{J}_\alpha(\bm{x},t)=\mathring{\lambda}\,h_\alpha(\bm{x},t)$ for
$\bm{x}\notin\mathfrak{B}$ and corresponding boundary terms $\propto
h_\alpha^{\mathfrak{B}}$. To explain the consequences, let us
introduce the generating functional
\begin{eqnarray}\label{eq:Wgenfctdef}
\mathcal{G}[\tilde{\bm{J}},\tilde{\bm{K}};\bm{J},\bm{K}]
&=&
\ln \bigg\langle\exp\bigg\{\int dt\,\bigg[\int_{\mathfrak{V}}
\big(\tilde{\bm{J}}\cdot\tilde{\bm{\phi}}+
\bm{J}\cdot\bm{\phi}\big)\nonumber\\ &&\strut +\int_{\mathfrak{B}}
\big(\tilde{\bm{K}}\cdot\tilde{\bm{\phi}}+
\bm{K}\cdot\bm{\phi}\big)\bigg]\bigg\}\bigg\rangle\;,
\end{eqnarray}
where $\tilde{\bm{K}}$ and $\bm{K}$ are source functions localized on
the boundary $\mathfrak{B}$. They serve to generate the boundary
operators
$\tilde{\bm{\phi}}^{\mathfrak{B}}=\tilde{\bm{\phi}}(\bm{x}_{\mathfrak{B}},t)$
and $\bm{\phi}^{\mathfrak{B}}=\bm{\phi}(\bm{x}_{\mathfrak{B}},t)$
with $\bm{x}_{\mathfrak{B}}\in\mathfrak{B}$ by functional
differentiation. In the case of periodic boundary conditions, they are
not needed, and we write $\mathcal{G}[\tilde{\bm{J}};\bm{J}]$ for the
analog of functional~\eqref{eq:Wgenfctdef}.

To specify a boundary point $\bm{x}_{\mathfrak{B}}$, we must say on
which boundary plane $\mathfrak{B}_j$ it is located and give its
lateral coordinate $\bm{r}$. Denoting the restriction of $\phi_\alpha$f
to $\mathfrak{B}_j$ by $\phi_\alpha^{\mathfrak{B}_j}$, we can write
the boundary operators as $\phi_\alpha^{\mathfrak{B}_j}(\bm{r},t)$ and
$\tilde{\phi}_\alpha^{\mathfrak{B}_j}(\bm{r},t)$. Whenever we do not
wish to specify on which surface plane these boundary operators are
localized, we continue writing $\phi_\alpha^{\mathfrak{B}}$ and
$\tilde{\phi}_\alpha^{\mathfrak{B}}$. Analogous conventions will be
used for the boundary sources $\tilde{K}^{\mathfrak{B}_j}_\alpha$ and
$K^{\mathfrak{B}_j}_\alpha$ and the boundary magnetic fields
$\mathring{h}_\alpha^{\mathfrak{B}_j}$.

Functional~\eqref{eq:Wgenfctdef} generates the cumulants
\begin{eqnarray}
  \label{eq:Wdef}
\lefteqn{\left\langle \prod_{i=1}^{\tilde{N}}
  \tilde{\phi}_{\tilde{\alpha}_i}\prod_{k=1}^{\tilde{M}}
  \tilde{\phi}_{\tilde{\beta}_k}^{\mathfrak{B}} \prod_{l=1}^{N}
  \phi_{\alpha_l} \prod_{m=1}^{M}\phi_{\beta_m}^{\mathfrak{B}}
\right\rangle^{\text{cum}}}\hspace{5em}&&\nonumber\\
&\equiv& \mathring{\lambda}^{-\tilde{N}-\tilde{M}}\,W^{(\tilde{N},\tilde{M};N,M)}
\end{eqnarray}
whose $\tilde{N}+N+\tilde{M}+M$ position vectors, time arguments, and
tensorial indices we have suppressed. From the correspondences
\begin{equation}
  \label{eq:hphitcorr}
  \frac{\mathring{\lambda}\,\delta}{\delta \tilde{J}_\alpha(\bm{x},t)}\leftrightarrow \frac{\delta
  }{\delta \mathring{h}_\alpha(\bm{x},t)}\leftrightarrow\mathring{\lambda}\,\tilde{\phi}_\alpha(\bm{x},t)
\end{equation}
and
\begin{equation}
  \label{eq:hbphitcorr}
   \frac{\mathring{\lambda}\,\delta}{\delta \tilde{K}^{\mathfrak{B}_j}_\alpha(\bm{r},t)}
   \leftrightarrow \frac{\delta 
  }{\delta \mathring{h}^{\mathfrak{B}_j}_\alpha(\bm{r},t)}\leftrightarrow\mathring{\lambda}\,
  \tilde{\phi}^{\mathfrak{B}_j}_\alpha(\bm{r},t) 
\end{equation}
it is clear that the functions $W^{(\tilde{N},\tilde{M};N,M)}$ defined
in Eq.~\eqref{eq:Wdef} are the  usual connected correlation and
response functions.

\subsection{Fluctuation-dissipation theorem and mesoscopic boundary conditions}
\label{sec:flucbc}

Several other remarks are in order here, which concern the
fluctuation-dissipation theorem and the boundary conditions 
on the mesoscopic scale where our continuum approximation applies.

First, the fluctuation-dissipation theorem\cite{HH77}
\begin{eqnarray}
  \label{eq:FDT}
  \lefteqn{\mathring{\lambda}\,
    \langle\phi_\alpha(\bm{x},t)\,
    \tilde{\phi}_{\alpha'}(\bm{x}',t')\rangle}&& \nonumber \\ &=& 
  -\theta(t-t')\,\partial_t\langle\phi_\alpha(\bm{x},t)
  \,\phi_{\alpha'}(\bm{x}',t')\rangle^{\text{cum}}  
\end{eqnarray}
holds. Second, the boundary contributions to the classical equations
of motions yield the boundary conditions
    \begin{eqnarray}
       \partial_n\tilde{\bm{\phi}}(\bm{x},t)&=&
  \mathring{c}_j\,\tilde{\bm{\phi}}(\bm{x},t)\,,
  \;\;\bm{x}\in\mathfrak{B}_j\,,\nonumber \\[\smallskipamount]
  \partial_n\bm{\phi}(\bm{x},t)&=&\mathring{c}_j\,\bm{\phi}(\bm{x},t)\,,
\;\;\bm{x}\in\mathfrak{B}_j\,.\label{eq:bc}
\end{eqnarray}
These hold beyond the classical approximation inside of averages (up
to anomalies at coinciding
points).\cite{DJ92,Die94b,WD95} On the level of the
classical (zero-loop) approximation, they ensure that the matrix
integral kernel
\begin{equation}
  \label{eq:J2}
  \bm{\mathcal{J}}^{(2)}\equiv
  \begin{pmatrix}
    2\mathring{\lambda}&\partial_t+\mathring{\lambda}(-\triangle+\mathring{\tau})\\
-\partial_t+\mathring{\lambda}(-\triangle+\mathring{\tau})&0
  \end{pmatrix}\otimes (\delta_{\alpha\beta})
\end{equation}
of the quadratic part of $\mathcal{J}[\tilde{\bm{\phi}},\bm{\phi}]$ is
self-adjoint.
Third, the boundary conditions \eqref{eq:bc} imply that the
fluctuation-dissipation theorem~\eqref{eq:FDT} remains valid when either
$\bm{x}$ or $\bm{x}'$ approaches a surface point. 

In the case of periodic boundary conditions, we have
\begin{eqnarray}\label{eq:pbc}
\tilde{\bm{\phi}}(\bm{x}+L\bm{e}_z,t)&=& \tilde{\bm{\phi}}(\bm{x},t)\,,
\nonumber\\[\smallskipamount] 
\bm{\phi}(\bm{x}+L\bm{e}_z,t)&=&\bm{\phi}(\bm{x},t)\,,
\end{eqnarray}
instead of Eq.~\eqref{eq:bc}.

\section{Perturbation theory and RG}
\label{sec:PRRen}

\subsection{Free propagators}
 \label{sec:pertth}
 To set up perturbation theory, we employ dimensional regularization
 and focus on the disordered phase. The free response and correlation
 propagators, $R_L$ and $C_L$, then follow from the inverse of the matrix
 kernel~\eqref{eq:J2}. We have
\begin{equation}
  \label{eq:freRC}
  \Big[\bm{\mathcal{J}}^{(2)}\Big]^{-1}\equiv \bm{G}_L=
  \begin{pmatrix}
    0&R_L^\dagger\\ R_L&C_L
  \end{pmatrix}\otimes (\delta_{\alpha\beta})\;.
\end{equation}
where $R_L$ is the solution to
\begin{equation}
  \label{eq:Rfree}
  \big[\partial_t+\mathring{\lambda}(-\triangle+\mathring{\tau})\big]\,R_L(\bm{x},t;\bm{x}',t')
  =\delta(\bm{x}-\bm{x}')\,\delta(t-t') \,,
\end{equation}
while $C_L$ is proportional to the convolution $R_L\ast R_L^\dagger$:
\begin{eqnarray}
  \label{eq:Cfree}
 \lefteqn{C_L(\bm{x},t;\bm{x}',t')=2\mathring{\lambda}\,(R_L\ast
 R_L^\dagger)(\bm{x},t;\bm{x}',t')}&&\nonumber\\ 
&=&2\mathring{\lambda} \int_{-\infty}^\infty
d\tilde{t}\int_{\mathfrak{V}}d^d\tilde{x}\,
R_L(\bm{x},t;\tilde{\bm{x}},\tilde{t})\, 
R_L(\tilde{\bm{x}},\tilde{t};\bm{x}',t') \,.\quad
\end{eqnarray}

The $\bm{y}t$-Fourier transforms of these quantities (for which we use
the notational conventions summarized in Appendix~\ref{sec:conv}) can
be expressed as 
\begin{equation}
  \label{eq:Rpzomega}
  R_L(\bm{p};z,z';\omega)=\sum_m\frac{f_m(z)\,f^*_m(z')}{-i\omega
    +\mathring{\lambda}\,(\mathring{\tau}+p^2+k_m^2)}
\end{equation}
and
\begin{equation}\label{eq:Cpzomega}
C_L(\bm{p};z,z';\omega)=\sum_m\frac{2\mathring{\lambda}\,f_m(z)\,f^*_m(z')}{\big|-i\omega
    +\mathring{\lambda}\,(\mathring{\tau}+p^2+k_m^2)\big|^2}\phantom{(2.19}
\end{equation}
in terms of a complete set of orthonormal eigenfunctions $f_m(z)$ of
the operator $-\partial_z^2$, where $f_m^*(z')$ is the complex conjugate of $f_m(z')$. These functions are properly normalized
solutions to
\begin{equation}
  \label{eq:fmdef}
  -\partial_z^2\,f_m(z)=k_m^2\,f_m(z)\;,
\end{equation}
subject to the boundary conditions
\begin{eqnarray}
  \label{eq:fmbc}
  f_m'(0)&=&\mathring{c}_1\,f_m(0)\;,\nonumber\\
-f_m'(L)&=&\mathring{c}_2\,f_m(L)\;.
\end{eqnarray}

For non-negative values of $\mathring{c}_1$ and $\mathring{c}_2$, the spectrum
$\{k^2_m\}$ is discrete with $k_m^2\ge 0$. The
eigenfunctions are phase-shifted cosine functions $f_m(z)=A_m
\cos(k_mz+\vartheta_m)$, whose phase shift $\vartheta_m$ follows from
the first of the boundary conditions~\eqref{eq:fmbc}. The eigenvalues
are solutions to the transcendental equation implied by the boundary
condition at $z=L$. For general values of $\mathring{c}_j\ge 0$, the
eigenvalues $k_m^2$ depend on both $\mathring{c}_1$ and $\mathring{c}_2$, as do the
normalization factors $A_m$ and the phase shifts $\vartheta_m$ (via
$k_m$).\cite{RS02,Sch08,SD08} For the special values
$(\mathring{c}_1,\mathring{c}_2)=(\infty,\infty)$, $(0,0)$, and $(\infty,0)$
corresponding to the combinations D-D, N-N, and D-N of Dirichlet (D)
and Neumann (N) boundary conditions on the two planes, the eigenvalues
$k_m^2$ and eigenfunctions $f_m$ can be found in Appendix A of
Ref.~\onlinecite{KD92a} and Appendix A of Ref.~\onlinecite{Kre94}.

However, the response propagator $R_L(\bm{p};z,z';\omega)$ can also be
determined by solving the analog of Eq.~\eqref{eq:Rfree} in the
$\bm{p}z\omega$~representation using familiar methods for
Sturm-Liouville differential equations.\cite{CH68} We give the result
of such a calculation for general non-negative values of $\mathring{c}_1$,
$\mathring{c}_2$, and $\mathring{\tau}$ in Eq.~\eqref{eq:Gfree} of Appendix~\ref{sec:frp}.
In the special cases $\mathring{c}_1=\mathring{c}_2=0$ and  $\mathring{c}_1=\mathring{c}_2=\infty$, the
result is equivalent to the representations\cite{Sym81,Die86a,GD06b}
\begin{eqnarray}
  \label{eq:RLDDNNir}
      R_L^{(\substack{\text{N-N}\\ \text{D-D}})}(\bm{p};z_1,z_2;\omega)&=& \sum_{m=-\infty}^\infty\big[
  R_\infty(\bm{p},z_1-z_2-m2L,\omega)\nonumber\\ &&\strut \pm
  R_\infty(\bm{p},z_1+z_2-m2L,\omega)\big]
\end{eqnarray}
of the corresponding Neumann and Dirichlet propagators
$R_L^{(\text{N-N})}$ and $R_L^{(\text{D-D})}$ as a sum of image
contributions involving the bulk propagator (see, e.g.,
Refs.~\onlinecite{Die86a}, \onlinecite{Sym81}, and
\onlinecite{GD08})
\begin{equation}
  \label{eq:R_b}
  R_\infty(\bm{p},z_{12},\omega)=\frac{1}{2\mathring{\lambda}\,\mathring{\kappa}_\omega}\,e^{-\mathring{\kappa}_\omega\,|z_{12}|}\;,
\end{equation}
where $z_{12}\equiv z_1-z_2$ and 
\begin{equation}
  \label{eq:kapom}
  \mathring{\kappa}_\omega=\sqrt{p^2+\mathring{\tau}-i\omega/\mathring{\lambda}}\;.
\end{equation} 

In the case of periodic boundary conditions, one has
\begin{eqnarray}
  \label{eq:RLperir}
  R_L^{(\text{per})}(\bm{p};z_1,z_2;\omega)= \sum_{m=-\infty}^\infty
  R_\infty(\bm{p},z_{12}-mL,\omega)\,.\;\;\;
\end{eqnarray}

The corresponding finite-size correlation propagators
$C_L^{(\text{N-N})}$, $C_L^{(\text{D-D})}$, and $C_L^{(\text{per})}$
can be expressed in terms of the free bulk correlation
propagator $C_\infty(\bm{p},z_{12},\omega)$ in a manner completely
analogous to Eqs.~\eqref{eq:RLDDNNir} and \eqref{eq:RLperir}.

\subsection{Reparametrizations}
\label{sec:ren}

As is well known and explained
elsewhere,\cite{Sym81,DD83b,Die86a,Die87,DJ92,Die94b}
the singularities of $R_\infty(\bm{x}_{12},t_{12})$ and
$C_\infty(\bm{x}_{12},t_{12})$ at coinciding points
$(\bm{x}_{12},t_{12})=(\bm{0},0)$ produce ultraviolet (uv)
singularities in Feynman integrals of the multipoint cumulant and
response functions~\eqref{eq:Wdef}. For dimensions $d\le 4$ and
periodic boundary conditions, the uv singularities of these functions
can be absorbed via standard ``bulk'' reparametrizations of the form
\begin{subequations}\label{zconstants}
\begin{eqnarray}
\bm{\phi}&=&Z_\phi^{1/2}\,\bm{\phi}_R,\\
\tilde{\bm{\phi}}&=&Z_{\tilde{\phi}}^{1/2}\,\tilde{\bm{\phi}}_R,  \\
\mathring{\lambda}&=&\mu^{-2}\,[Z_\phi/Z_{\tilde{\phi}}]^{1/2}\lambda,  \\
\delta\mathring{\tau}\equiv \mathring{\tau}-\mathring{\tau}_c&=&\mu^2\,Z_\tau\tau,  \\
\mathring{u}\,N_d&=&\mu^\epsilon Z_uu.
\end{eqnarray}
Here $\mu$ is an arbitrary momentum scale and $\mathring{\tau}_{c}$ denotes the
critical value of $\mathring{\tau}$ of the $d$-dimensional bulk theory. Following
Ref.~\onlinecite{GD08}, we choose the factor that is absorbed in the renormalized
coupling constant $u$ as
\begin{eqnarray}
  \label{eq:Nd}
  N_d&=&\frac{2\,\Gamma(3-d/2)}{(d-2)(4\pi)^{d/2}}\nonumber \\ &=&
\frac{1}{16\pi^2}\left[1+\frac{1-\gamma_E+\ln(4\pi)}{2}\,\epsilon
  +O(\epsilon^2)\right], \phantom{(2.15)}
\end{eqnarray} 
\end{subequations}
where $\gamma_E=-\Gamma'(1)$ is Euler's constant. If we employ
dimensional regularization and fix the renormalization factors $Z_g$,
$g=\phi,\tilde{\phi},u,\tau$, by minimal subtraction of poles in
$\epsilon$, this convention ensures that the two-loop results for
these functions given in equations (3.42a)--(3.42c) and (4.54) of
Ref.~\onlinecite{Die86a} apply.

For free boundary conditions, additional primitive uv singularities
with support on $\mathfrak{B}_1$ and $\mathfrak{B}_2$ occur. These can
be absorbed through the additional (``surface'') reparametrizations
\begin{eqnarray}\label{eq:surfreps}
    \delta\mathring{c}_j&\equiv&\mathring{c}_j
    -\mathring{c}_{\mathrm{sp}}=\mu Z_c\,c_j,\nonumber\\ 
    \bm{\phi}^{\mathfrak{B}}&=&(Z_\phi\, Z_1)^{1/2}\,
    [\bm{\phi}^{\mathfrak{B}}]_R,\nonumber\\
 \tilde{\bm{\phi}}^{\mathfrak{B}}&=&(Z_{\tilde{\phi}}\, Z_1)^{1/2}\,
    [\tilde{\bm{\phi}}^{\mathfrak{B}}]_R,
\end{eqnarray}
known from the semi-infinite case, where
$[\bm{\phi}^{\mathfrak{B}}]_R$ and
$[\tilde{\bm{\phi}}^{\mathfrak{B}}]_R$ are renormalized boundary
operators. Explicit two-loop expressions for the renormalization
factors $Z_1$ and $Z_c$ may be found in Eqs.~(3.66a) and (3.66b) of
Ref.~\onlinecite{Die86a} or in Refs.~\onlinecite{DD81b} and \onlinecite{DD83a}.

\subsection{RG equations and scaling} \label{sec:RGE} 

Upon introducing the renormalized functions
\begin{eqnarray}
  \lefteqn{W^{(\tilde{N},\tilde{M};N,M)}_R }&&\nonumber\\ &=& Z_\phi^{-(\tilde{N}+N)/2} (Z_\phi
Z_1)^{-(\tilde{M}+M)/2}\, W^{(\tilde{N},\tilde{M};N,M)}\,,\phantom{3.14}
\end{eqnarray}
we can
exploit the invariance of the bare functions
$W^{(\tilde{N},\tilde{M};N,M)}$ under changes $\mu\to\mu\ell$ in a
standard fashion to obtain the RG equations
\begin{equation}\label{eq:RGE}
\bigg[\mathcal{D}_\mu
+(\tilde{N}+N)\,\frac{\eta_{\phi}}{2}
+(\tilde{M}+M)\,\frac{\eta_\phi+\eta_1}{2}\bigg] 
W^{(\tilde{N},\tilde{M};N,M)}_R=0\;.
\end{equation}
Here
\begin{equation}
  \label{eq:Dmu}
\mathcal{D}_\mu=\mu\partial_\mu+\sum_{g=u,\tau,\lambda,c_1,c_2}\beta_g
\,\partial_g \;.
\end{equation}
The beta and exponent functions $\beta_g$ and $\eta_g$ are given by
\begin{eqnarray}
  \label{eq:betadef}
  \beta_g&\equiv &\mu\partial_\mu\vert_0 g=-(d_g+\eta_g)g\nonumber\\
&=&
\begin{cases}
  -(\epsilon+\eta_u)u\,,&g=u\,,\\
-(2+\eta_\tau)\tau\,,&g=\tau\,,\\
(2-\eta_\lambda)\lambda\,,&g=\lambda\,,\\
-(1+\eta_c)c_j\,,&g=c_j\,,\;j=1,2\,.
\end{cases}
\end{eqnarray}
and
\begin{equation}
  \label{eq:etawdef}
  \eta_g\equiv\mu\partial_\mu|_0g\;,\quad g=u,\tau,\phi,\tilde{\phi},\lambda,1,c_1,c_2\,,
\end{equation}
respectively, where $\partial_\mu|_0$ means a derivative at fixed
parameters of the bare theory. Explicit results for $\beta_u$ to order
$u^3$ and for $\eta_\phi(u)$, $\eta_\tau(u)$, $\eta_1(u)$, and
$\eta_c(u)\equiv\eta_{c_j}(u)$ to order $u^2$ may be looked up in
Eqs.~(3.75a), (3.75b), (3.76a), and (3.76b) of Ref.~\onlinecite{Die86a},
respectively. The function $\eta_\lambda$, which can be written as
$\eta_\lambda=(\eta_\phi-\eta_{\tilde{\phi}})/2$, is given to order
$u^2$ in the first line of Eq.~(III.8) of
Ref.~\onlinecite{DD83b}.

It should be obvious how the above RG equations carry over to the case
of periodic boundary conditions: the RG equation for
$W^{(\tilde{N};N)}_R$ agrees with Eq.~(\ref{eq:RGE}) if we set  
$\tilde{M}=M=0$, except that the contributions to $\mathcal{D}_\mu$
involving $\partial_{c_j}$ drop out because these variables do not
occur in the theory when periodic boundary conditions are chosen.

Using characteristics the RG equations~\eqref{eq:RGE} can be exploited
in a familiar fashion to derive the asymptotic scaling behavior of the
functions $W^{(\tilde{N},\tilde{M};N,M)}_R$. Let $\bar{g}(\ell)$ be
solutions to the flow equations
\begin{equation}
  \label{eq:flowdeg}
  \ell\, \frac{d\bar{g}(\ell)}{d\ell}=\beta_g[\bar{u}(\ell),\bar{g}(\ell)]\,,\quad g=u,\tau,\lambda,c_1,c_2,
\end{equation}
satisfying the initial conditions $\bar{g}(1)=g$. Then we have
\begin{equation}
  \label{eq:ubarsol}
  \ln\ell=\int_u^{\bar{u}}\frac{dv}{\beta_u(v)}
\end{equation}
and
\begin{equation}
  \label{eq:gflow}
  \bar{g}(\ell)=E_g[\bar{u}(\ell),u]\,\ell^{-(d_g+\eta_g^*)}\,g,
  \quad g=\tau,\lambda,c_1,c_2\,,  
\end{equation}
where $E_g(\bar{u},u)$ are the trajectory integrals
\begin{equation}
  \label{eq:Eg}
  E_g(\bar{u},u)=\exp\left[-\int_u^{\bar{u}}dv\,
    \frac{\eta_g(v)-\eta_g^*}{\beta_u(v)}\right] 
\end{equation}
and the asterisk indicates values $\eta_g^*\equiv\eta_g(u^*)$ at the
nontrivial root $u^*=O(\epsilon)$ of $\beta_u(u)$.

To illustrate the consequences of the RG equations~\eqref{eq:RGE}, let
us consider the multipoint cumulants in the $\bm{x}t$-presentation.
Writing $W^{(\tilde{N},\tilde{M};N,M)}_R\equiv
W_R(\bm{x},t;u,\tau,c_1,c_2,L,\lambda,\mu)$, where $\bm{x}$ and $t$
represent all position and time variables, respectively, we note
that the dimension of this function is
$\lambda^{\tilde{N}+\tilde{M}}\, \mu^{d_W}$, with 
\begin{equation}
  \label{eq:dW}
  d_W=(\tilde{N}+\tilde{M})\,\frac{d+2}{2}+(N+M)\,\frac{d-2}{2}\;.
\end{equation}
Using this in conjunction with the solution to the RG
equation~\eqref{eq:RGE}, one arrives at
\begin{eqnarray}
  \label{eq:Wscal}
\lefteqn{W_R(\bm{x},t;u,\tau,c_1,c_2,L,\lambda,\mu)=
  \lambda^{\tilde{N}+\tilde{M}}\, 
  \mu^{d_W}\,\ell^{\Delta_W}}&&\nonumber\\[\smallskipamount] &&
 \times E_W(\ell)\,
 W_R(\mu\ell\bm{x},\bar{\lambda}\mu^2t;\bar{u},\bar{\tau},
 \bar{c}_1,\bar{c}_2,\mu \ell L,1,1)\,.\qquad  
\end{eqnarray}
Here $\Delta_W$ is the scaling dimension of $W_R$. We use the
notation $\Delta[\mathcal{O}]$ for the scaling dimension
of an operator $\mathcal{O}(\bm{x},t)$. Then we have
\begin{eqnarray}
  \label{eq:scdim}
  \Delta[\phi]&\equiv& (d-2+\eta_\phi^*)/2=\beta/\nu\,,\nonumber
  \\[\smallskipamount] 
 \Delta[\tilde{\phi}]&\equiv&
 2-\eta_\lambda^*+\Delta[\phi]=\mathfrak{z}+\beta/\nu\,,\nonumber
 \\[\smallskipamount] 
 \Delta[\phi^{\mathfrak{B}}]&\equiv& (d-2+\eta_\phi^*
 +\eta_1^*)/2=\beta_1^{\text{sp}}/\nu\,,\nonumber\\[\smallskipamount] 
 \Delta[\tilde{\phi}^{\mathfrak{B}}]&\equiv& \mathfrak{z}
 +\Delta[\phi^{\mathfrak{B}}]=\mathfrak{z} +\beta_1^{\text{sp}}/\nu\,,
\end{eqnarray}
where we have introduced the standard static bulk critical indices
$\beta$ and $\nu$, the dynamic bulk critical exponent $\mathfrak{z}$,
and the surface critical exponent $\beta_1^{\text{sp}}$ of the
special transition.\cite{Die86a} In terms of these quantities, the
scaling dimension reads
\begin{equation}
  \label{eq:DelW}
  \Delta_W= \tilde{N}\Delta[\tilde{\phi}] +N\Delta[\phi]
  +\tilde{M}\Delta[\tilde{\phi}^{\mathfrak{B}}]+
  M\Delta[\phi^{\mathfrak{B}}]\,.
\end{equation}
Further, the prefactor $E_W(\ell)$ denotes the trajectory integral
\begin{equation}
  \label{eq:EW}
  E_W(\ell)=\exp\bigg\{\int_u^{\bar{u}(\ell)}
  \frac{\eta_W(v)-\eta_W^*}{\beta_u(v)}\,dv\bigg\} 
\end{equation}
with
\begin{equation}
  \label{eq:etaW}
  \eta_W(u)=\frac{\tilde{N}+N}{2}\,\eta_\phi
  +\frac{\tilde{M}+M}{2}\,(\eta_\phi+\eta_1)
  -(\tilde{N}+\tilde{M})\,\eta_\lambda\,.  
\end{equation}

It is easily checked that Eq.~\eqref{eq:Wscal} yields scaling forms in
accordance with the phenomenological theory of finite-size
scaling.\cite{Bar83,Pri90}  Assuming that the initial
coupling constant $u$ is nonzero, we replace the running coupling
constant $\bar{u}$ and the scale-dependent amplitude $E_W(\ell)$ by
their respective long-scale limits $\bar{u}(0)=u^*$ and $E_W^*=E_W(0)$ and substitute
the running variables $\bar{\tau}$, $\bar{\lambda}$, and $\bar{c}_j$ by
their limiting forms
\begin{equation}
  \label{eq:gbaras}
  \bar{g}(\ell) \underset{\ell\to 0}{\approx}
  E_g^*(u)\,\ell^{-(d_g+\eta_g^*)}\,g =
 \begin{cases}
     E_\tau^*(u)\,\ell^{-1/\nu}\,\tau,\\
 E_\lambda^*(u)\,\ell^{\,\mathfrak{z}}\,\lambda,\\
 E_{c}^*(u)\,\ell^{-\Phi/\nu}\,c_j,
 \end{cases}
   \end{equation}
   where $E_g^*(u)\equiv E_g(u^*,u)$ are nonuniversal amplitudes, and
   we have introduced the surface crossover exponent
   $\Phi=\nu(1+\eta_c^*)$.\cite{Die86a}  If we now fix the scale
   parameter $\ell\equiv \ell_\tau$ for given $\tau\ne 0$ by 
   $\bar{\tau}(\ell_\tau)=\text{sign}(\tau)$, then
\begin{equation}
  \label{eq:xiinf}
 \mu^{-1}\ell_\tau^{-1}\underset{\tau\to \pm 0}{\approx}
 \mu^{-1}\,\big(E_\tau^*|\tau|\big)^{-\nu}
\end{equation}
agrees with the second-moment bulk
correlation length  
\begin{equation}
  \label{eq:xipm}
  \xi_\infty^\pm\approx \xi^{(0)}_\pm\,|T/T_{c,\infty}-1|^{-\nu}\;,
\end{equation}
up to a nonuniversal scale factor. We thus obtain the scaling forms
 \begin{eqnarray}
   \label{eq:Wscf}
\lefteqn{W_R(\bm{x},t;u,\tau,c_1,c_2,L,\lambda,\mu)\approx \lambda^{\tilde{N}+\tilde{M}}\, \mu^{d_W}}&&\nonumber\\ &&\times
 \,  (\mu\xi_\infty^\pm)^{-\Delta_W}E_W^*\,
 \Xi_\pm(\bm{x}/\xi_\infty^\pm,\mathsf{t};
 \mathsf{c}_1,\mathsf{c}_2,L/\xi_\infty^\pm)\phantom{(3.15)}\;,
 \end{eqnarray}
in which $\mathsf{t}$ represents the set of all scaled time variables
\begin{equation}
  \label{eq:tscal}
  \mathsf{t}=E_\lambda^*\lambda\mu^2t\,
  (\mu\xi_\infty^\pm)^{-\mathfrak{z}} \,,
\end{equation}
while $\mathsf{c}_1$ and $\mathsf{c}_2$ denote the scaled surface
variables
\begin{equation}
   \label{eq:cscal}
   \mathsf{c}_j= E_c^*c_j\,(\mu\xi_\infty^\pm)^\Phi\,. 
\end{equation}
The scaling functions $\Xi_\pm$ are given by the restrictions of
$W_R(\dotsc;\bar{u},\bar{\tau},\dotsc)$ to the hyperplanes with
$\bar{u}=u^*$ and $\bar{\tau}=\pm 1$.

Analogous results apply for periodic boundary conditions,
except that the surface scaling variables $\mathsf{c}_1$ and
$\mathsf{c}_2$ are missing.

\section{RG-improved perturbation theory}
\label{sec:RGipt}

\subsection{Background}
\label{sec:bg}

There exist known important cases of critical phenomena, in which the
infrared (IR) singularities one is concerned with originate from the divergence
of a single length, such as the bulk correlation length $\xi_\infty$.
The most familiar example is static critical behavior that occurs at a
usual bulk critical point when it is approached from the disordered
phase. The crux of RG-improved perturbation theory is to exploit the
RG to relate the behavior of systems with large $\xi_\infty$ to that
of systems with $\xi_\infty$ of order unity and then determine
properties of the latter by means of appropriate perturbation theory
techniques. Provided no other sources of IR singularities
exist, the resulting perturbation expansions will not be plagued by
  problems even though sophisticated resummation methods may well be
required to obtain reliable results.\cite{ZJ02}

In order to determine the behavior of properties at the critical
point, information about the behavior of the static bulk analogs of
the above scaling functions $\Xi_\pm$ in the limit
$\bm{x}/\xi_\infty\to \bm{0}$ is required. In the simple case
considered so far, this can be inferred from the requirement
that a temperature independent nonzero limit results for the property
considered. Furthermore, the form of corrections to the leading
asymptotic behavior of the pair correlation function for
$|\bm{x}|/\xi_\infty\ll 1$ can be determined via the short-distance
expansion (see, e.g., Refs.~\onlinecite{AF74}, \onlinecite{AM-M05},
and their references).

However, even in the case of static bulk critical behavior, additional
sources of IR singularities may exist. For example, in systems
exhibiting spontaneous breakdown of continuous symmetries, IR
singularities associated with Goldstone modes appear on the
coexistence curve.\cite{SH78} Other examples are provided by crossover
phenomena such as the crossover away from a bicritical point. Here the
crossover from one type of critical behavior to another leads to
singularities in the corresponding crossover scaling
functions.\cite{AG78,AM-M05} To ensure that these IR singularities
are properly taken into account in RG-improved perturbation theory is
a nontrivial task. For some cases, specially designed RG schemes
exists that allow one to correctly build in the asymptotic behavior at
the stable fixed point to which the crossover occurs.\cite{AF74,Law81}
Within the framework of dimensionality expansions, this usually works
when the additional IR singularities are properly handled by simple
theories such as the random phase approximation\cite{SH78,Law81} or
are accessible to an $\epsilon=d^*-d$ expansion about the same upper
critical dimension $d^*$.\cite{AG78,AM-M05} Much more challenging
are problems involving \emph{two distinct kinds of nontrivial critical
  behavior} with \emph{different upper critical dimensions}.

Unfortunately, the study of critical behavior in films of finite
thickness belongs to the latter class of hard problems. The reason is
that a full treatment would involve a proper analysis of
\emph{dimensional crossover}. Suppose the film undergoes for finite
$L$ and given bulk dimension $d$ a sharp phase transition at a
temperature $T_{c,L}$. If all interactions are ferromagnetic, this
temperature must satisfy $T_{c,L}\le T_\infty$ by
Griffiths-Kelly-Sherman inequalities.\cite{Gri68,KS68} On lowering the
temperature from an initial temperature $T>T_{c,\infty}$, one will
therefore first observe $d$-dimensional critical behavior as $L,
\xi_\infty\to \infty$ with $L/\xi_\infty\gg 1$, which will cross over
to $(d-1)$-dimensional critical behavior as the length $\xi_\|$ on
which correlations along the film decay becomes much larger than $L$.
\cite{Bar83} Since $d$- and $(d-1)$-dimensional critical behaviors
involve different upper critical dimensions ($d^*=4$ and $d^*=5$,
respectively), one cannot handle both of them by the same
dimensionality expansion. In those cases where no sharp phase
transition is possible for finite $L$, a rounded transition will occur
at $T_{c,L}$. Even then, the $\epsilon$ expansion must not be expected
to correctly capture the behavior for $T\simeq T_{c,L}$.
Because of these and technical difficulties, previous investigations
of static finite-size effects in films based on the $\epsilon$
expansion\cite{KD92a,KD91,KED95,fbkrech,Kre94,DGS06,GD08} have
focused on the case $T\ge T_{c,\infty}$, even though  RG-improved
mean-field results for $T<T_c$ exist for ($d=3$)-dimensional systems
with bulk critical and tricritical
points.\cite{ZSRKC07,MGD07,remHewett,GC99,Huc07,VGMD07}

For simplicity, we shall restrict ourselves here also to
the case $T\ge T_{c,\infty}$. We begin with a discussion of the
breakdown of the $\epsilon$ expansion at $T_{c,\infty}$ for periodic
and sp-sp boundary conditions. Since the latter is the only case of
free boundary conditions explicitly considered henceforth, we shall no
longer keep track of the enhancement variables, fixing $c_1$ and $c_2$
at zero whenever we deal with this boundary condition.

\subsection{Breakdown of $\bm{\epsilon}$ expansion at $\bm{T_{c,\infty}}$ and construction of effective action}
  \label{sec:breakdeps}

We shall use dimensional regularization. In a perturbative approach,
the critical values $\mathring{\tau}_c$ and $\mathring{c}_{\text{sp}}$ then
vanish.\cite{DD81b,DD83a,Die86a} In any case, the response and correlation
propagators for sp-sp boundary conditions satisfy Neumann boundary
conditions on both planes ($\wp=\text{N-N}$) at zero-loop order. 

The eigenvalues $k_m^2$ of $-\partial_z^2$ for periodic
($\wp=\text{per}$) and N-N boundary conditions are given by
$k^{(\mathrm{per})}_m=2\pi m/L$ with $m\in \mathbb{Z}$ and
$k^{(\mathrm{N-N})}_m=\pi m/L$ with $m=0,1,\dotsc,\infty$,
respectively.\cite{KD92a} In both cases, one has the
eigenvalue $k_0^2=0$. As can be seen from the mode
decompositions~\eqref{eq:Rpzomega} and \eqref{eq:Cpzomega}, the
associated $k_0=0$ modes give IR singular contributions to the free
propagators $R_L$ and $C_L$ at the bulk critical temperature ($\mathring{\tau}=0$)
when $\omega=p=0$. This is an artifact of the zero-loop approximation,
which predicts that a sharp phase transition occurs for finite $L$
precisely at $T_{c,\infty}$, yielding no shift $T_{c,L}-T_{c,\infty}$.
For bulk and semi-infinite systems the contributions from these zero
modes are negligible. However, for films of finite thickness $L$, we
must be more careful.

It is known from Refs.~\onlinecite{DGS06} and
\onlinecite{GD08} that beyond zero-loop order, the
${k=0}$~component of the order parameter becomes critical at the shifted
values $\delta\mathring{\tau}=-\delta\mathring{\tau}_L^{(\text{per})}$, and
$-\delta\mathring{\tau}_L^{(\text{sp-sp})}$ given by
\begin{equation}
  \label{eq:shifts}
   \delta\mathring{\tau}_L^{(\mathrm{per})}= 2^{2-\epsilon}\,
   \delta\mathring{\tau}_L^{\text{(sp-sp)}}
= \mathring{u}\,\frac{n+2}{6}\,
\frac{\Gamma(1-\epsilon/2)\, \zeta(2-\epsilon)}{2\pi^{2-\epsilon/2}\,
  L^{2-\epsilon}}\,,
\end{equation}
where $\zeta(s)$ denotes Riemann's zeta function. In order to obtain a
well-defined RG-improved perturbation theory at $T_{c,\infty}$, we
generalize the approach of Refs.~\onlinecite{DGS06} and
\onlinecite{GD08} to dynamics. Writing
\begin{eqnarray}
  \label{eq:phiphitsplit}
  \bm{\phi}(\bm{y},z,t)&=&L^{-1/2}\,\bm{\varphi}(\bm{y},t)+\bm{\psi}(\bm{y},z,t)\nonumber\\
\tilde{\bm{\phi}}(\bm{y},z,t)&=&L^{-1/2}\,\tilde{\bm{\varphi}}(\bm{y},t)+\tilde{\bm{\psi}}(\bm{y},z,t)
\end{eqnarray}
with
\begin{equation}
  \int_0^L\bm{\psi}(\bm{y},z,t)\,dz=\int_0^L\tilde{\bm{\psi}}(\bm{y},z,t)\,dz=\bm{0}\,,
\end{equation}
we split the fields $\bm{\phi}(\bm{x},t)$ and
$\tilde{\bm{\phi}}(\bm{x},t)$ into their zero-mode components
$\bm{\varphi}(\bm{y},t)$ and $\tilde{\bm{\varphi}}(\bm{y},t)$ and their
$k\ne 0$ orthogonal complements $\bm{\psi}(\bm{x},t)$ and
$\tilde{\bm{\psi}}(\bm{x},t)$. Upon integrating out the latter fields,
we  define an  effective action
$\mathcal{J}_{\text{eff}}[\tilde{\bm{\varphi}},\bm{\varphi}]$ for the
zero-mode components by
\begin{eqnarray}
  \label{eq:Jeffdef}
  \mathcal{J}_{\text{eff}}[\tilde{\bm{\varphi}},\bm{\varphi}] &\equiv
  &
  -\ln\text{Tr}_{\tilde{\bm{\psi}},\bm{\psi}}\,
  e^{-\mathcal{J}[\tilde{\bm{\psi}}+L^{-1/2}\tilde{\bm{\varphi}}, \bm{\psi}+L^{-1/2}\bm{\varphi}]}\nonumber\\
&=&\mathcal{J}^{[0]}_{\text{eff}}[\tilde{\bm{\varphi}},\bm{\varphi}\big]
-\ln\left\langle e^{-\mathcal{J}_{\text{int}}[ \tilde{\bm{\varphi}}, \tilde{\bm{\psi}},\bm{\varphi},\bm{\psi}]}
  \right\rangle_{\tilde{\bm{\psi}},\bm{\psi}}\,.\quad
\end{eqnarray}
where
\begin{equation}
  \label{eq:avpsitpsi}
  \langle\centerdot\rangle_{\tilde{\bm{\psi}},\bm{\psi}}
  =\int \centerdot
  \;e^{-\mathcal{J}[\tilde{\bm{\psi}},\bm{\psi}]}\,
  \mathcal{D}[\tilde{\bm{\psi}},\bm{\psi}] \,.
\end{equation}

Here
\begin{eqnarray}\label{eq:J0}
\mathcal{J}^{[0]}_{\text{eff}}[\tilde{\bm{\varphi}},\bm{\varphi}]&= &
\mathcal{J}\big[L^{-1/2}\tilde{\bm{\varphi}},L^{-1/2}\bm{\varphi}]
\nonumber\\ &=&
\int dt\int d^{d-1}y
\bigg\{\tilde{\bm{\varphi}}\cdot\bigg[\mathring{\lambda}\,
\bigg(\mathring{\tau}-\nabla_{\bm{y}}^2 
\nonumber\\ &&\strut +\frac{\mathring{u}}{6L}\,\varphi^2\bigg)\bm{\varphi} 
+\dot{\bm{\varphi}}-\mathring{\lambda}\,\tilde{\bm{\varphi}}\bigg]
\bigg\}
\end{eqnarray}
is the zero-loop contribution to
$\mathcal{J}_{\text{eff}}[\tilde{\bm{\varphi}},\bm{\varphi}]$. The
interaction part $\mathcal{J}_{\text{int}}$ is given by
\begin{eqnarray}
  \label{eq:Jint}
  \lefteqn{\mathcal{J}_{\text{int}}[\tilde{\bm{\varphi}},\tilde{\bm{\psi}},\bm{\varphi},\bm{\psi}]}&&\nonumber \\ &=&\frac{\mathring{\lambda}\mathring{u}}{6L}
\int dt\int_{\mathfrak{V}}
\Big[\psi^2\,(\tilde{\bm{\varphi}}\cdot\bm{\varphi})
 +2(\tilde{\bm{\varphi}}\cdot\bm{\psi}) (\bm{\varphi}
\cdot\bm{\psi})
\nonumber\\
&&\strut +\varphi^2\,(\tilde{\bm{\psi}}\cdot\bm{\psi}) +2(\tilde{\bm{\psi}}\cdot\bm{\varphi}) (\bm{\varphi}
\cdot\bm{\psi})
\nonumber\\ &&\strut
+L^{-1/2}(\tilde{\bm{\psi}}\cdot\bm{\varphi})\,\psi^2
+L^{-1/2}(\tilde{\bm{\varphi}}\cdot\bm{\psi})\,\psi^2  \Big]. 
\end{eqnarray}

Computing $\mathcal{J}_{\text{eff}}$ in a loop expansion and writing
\begin{equation}
  \label{eq:Jefflexp}
  \mathcal{J}_{\text{eff}}[\tilde{\bm{\varphi}},\bm{\varphi}]=
  \mathcal{J}^{[0]}_{\text{eff}}[\tilde{\bm{\varphi}},\bm{\varphi}]
  +\mathcal{J}_{\text{eff}}^{[1]}[\tilde{\bm{\varphi}},\bm{\varphi}] +
  \dotso\,,
\end{equation}
one easily derives the one-loop contribution
\begin{eqnarray}
  \label{eq:Jeff1}
\mathcal{J}_{\mathrm{eff}}^{[1]}[\tilde{\bm{\varphi}},\bm{\varphi}]&=&\frac{1}{2}\,\mathrm{Tr}
 \ln\Big[\openone +\frac{\mathring{\lambda}\mathring{u}}{L}\,\bm{G}^{(\wp)}_{L,\psi}
  \cdot\bm{B}\Big]
\end{eqnarray}
(cf.\ Ref.~\onlinecite{Die87}), where $\bm{G}^{(\wp)}_{L,\psi}$ denotes
the free propagator in the $k\ne0$~subspace for the respective
boundary conditions $\wp=\text{per}$ and $\text{sp-sp}$. Let
$P_0\equiv\vert k_0\rangle\langle k_0\vert$ be the projector onto
this $k=0$ space and $Q_0=\openone -P_0$. Then we may write
\begin{eqnarray}
  \label{eq:PLpsi}
  \lefteqn{\bm{G}_{L,\psi}=Q_0\bm{G}_LQ_0}&&\nonumber\\
 &=& \begin{pmatrix}
    0&R_{L,\psi}(\bm{y}_2,t_2;\bm{y}_1,t_1)\\
R_{L,\psi}(\bm{y}_1,t_1;\bm{y}_2,t_2)&C_{L,\psi}(\bm{y}_1,t_1;\bm{y}_2,t_2)
  \end{pmatrix}\otimes(\delta_{\alpha\beta})\,,\nonumber\\
\end{eqnarray}
where $R_{L,\psi}$ and $C_{L,\psi}$ are given by the analogs of
Eqs.~\eqref{eq:Rpzomega} and \eqref{eq:Cpzomega} one obtains by restricting their
summations over $m$ to values $m\ne 0$.

The operator $\bm{B}$ is block diagonal in
$\bm{y}t$~space. Introducing the familiar symmetric tensor
\begin{equation}
  \label{eq:Stens}
S_{\alpha\beta\gamma\delta}=\frac{1}{3}\,\left(\delta_{\alpha\beta}\delta_{\gamma\delta}
+\delta_{\alpha\gamma}\delta_{\beta\delta}
+\delta_{\alpha\delta}\delta_{\beta\gamma}\right) \,,
\end{equation}
we have
\begin{eqnarray}
  \label{eq:Bmat}
  \lefteqn{\bm{B}=\delta(\bm{y}_1-\bm{y}_2)\,\delta(t_1-t_2)}&&\nonumber\\
  &&\times\left.\begin{pmatrix}
    0&\frac{1}{2}\,
    S_{\alpha\beta\gamma\delta}\,\varphi_\gamma\,\varphi_\delta  \\[\medskipamount]
\frac{1}{2}\,
    S_{\alpha\beta\gamma\delta}\,\varphi_\gamma\,\varphi_\delta & 
    S_{\alpha\beta\gamma\delta}\,\tilde{\varphi}_\gamma\,\varphi_\delta
  \end{pmatrix}\right|_{\substack{\tilde{\bm{\varphi}}=\tilde{\bm{\varphi}}(\bm{y}_1,t_1)\\ \bm{\varphi}=\bm{\varphi}(\bm{y}_1,t_1)}}.\phantom{(4.12)}
\end{eqnarray}

\begin{figure}[htb]
  \centering
  \includegraphics[width=23em]{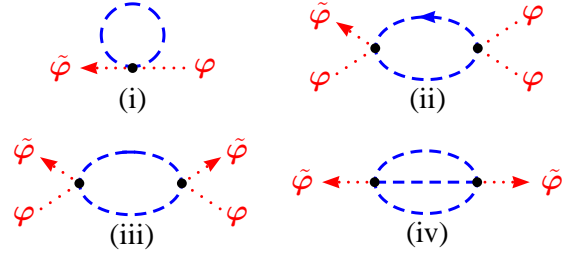}
  \caption{Graphs contributing to minus the effective
    action~\eqref{eq:Jefflexp}. Lines with and without an arrow
    represent the response and correlation propagators $R_L$ and
    $C_L$, respectively. Red dotted lines indicate $k=0$ components;
    blue broken lines $k\ne 0$ components (color online). For further
    explanations, see main text.}
  \label{fig:graphs}
\end{figure}

The graphs (i)--(ii) depicted in Fig.~\ref{fig:graphs}
are the contributions implied by $\mathcal{J}_{\text{eff}}^{[1]}$ that involve two
and four fields, respectively. Unlike graph (i), which is local in
$\bm{y}$ and $t$, graphs (ii)--(iv) depend on the
separations in $\bm{y}$ and $t$ between their two vertices (marked as
black dots) and thus are nonlocal. Similar nonlocal contributions
involving $6,8,\dotsc$ fields are contained in
$\mathcal{J}_{\text{eff}}^{[1]}[\tilde{\bm{\varphi}},\bm{\varphi}]$.
At one-loop order no modification of the Onsager coefficient $\mathring{\lambda}$
appears. However, the two-loop graph (iv) produces a nonlocal
modification of it.

Let us use the notations 
\begin{eqnarray}
  \label{eq:gamma11}
  \lefteqn{\gamma_{\tilde{\alpha}_1, \dotsc,\alpha_k}^{(\tilde{k};k)}
  (\tilde{\bm{y}_1}, 
  \tilde{t}_1;\dotsc;\bm{y}_k,t_k)}\nonumber\\ &=&
\frac{\delta^{\tilde{k}+k} \mathcal{J}_{\text{eff}}}{
  \delta\tilde{\varphi}_{\tilde{\alpha}_1}(\bm{y}_{\tilde{\alpha}_1},
  \tilde{t}_1) \dotsb
  \delta\varphi_{\alpha_k}(\bm{y}_k,t_k)}
\Big|_{\tilde{\bm{\varphi}}=\bm{\varphi} =\bm{0}}
\end{eqnarray}
for the effective vertices of $\mathcal{J}_{\text{eff}}$, and write
$\gamma^{(k)}_{\text{st}}(\bm{y}_1,t_1;\dotsc\bm{y}_k,t_k)$ for the
analogously defined static effective $k$-point vertex functions (which
were denoted by $\gamma^{(k)}$ in Ref.~\onlinecite{GD08}) .  By
analogy with the bulk case, the vertex functions
$\gamma^{(1;1)}(\dotsc;\omega)$ and
$\gamma^{(1;3)}(\dotsc;\{\omega_i\})$ must satisfy the relations
\begin{equation}
  \label{eq:gamma11gamma2stat}
  \gamma^{(1;1)}(\dotsc;\omega=0)=\mathring{\lambda}\,\gamma^{(2)}_{\text{st}}(\dotsc)
\end{equation}
and
\begin{equation}
  \label{eq:gamma13gamma4}
  \gamma^{(1;3)}(\dotsc;\{\omega_i=0\})=\mathring{\lambda}\,
  \gamma^{(4)}_{\text{st}}(\dotsc) \,,
\end{equation}
where the ellipses stand for the positions $\bm{y}_i$ or momenta
$\bm{p}_i$. The contribution of graph (i) to
$\gamma^{(1;1)}(\dotso;\omega)/\mathring{\lambda}$ is independent of $\omega$ and
hence agrees with that of its static analog
\raisebox{-0.5em}{\includegraphics[width=5em]{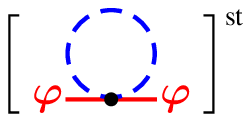}} to the
static two-point vertex function $\gamma^{(2)}_{\text{st}}$. It gives
an $L$-dependent shift of the temperature variable, changing it to
\begin{equation}
  \label{eq:tbL}
  \mathring{\tau}^{(\wp)}_L(\mathring{\tau})=\mathring{\tau}+\frac{n+2}{6}\,\mathring{u}\,I_1^{(\wp)}(L;\mathring{\tau})\;,
\end{equation}
where  the integrals
\begin{equation}
  \label{eq:I1wpdef}
  I_1^{(\wp)}(L;\mathring{\tau})=\sum_{m\ne 0}\int_{\bm{p}} \int_0^L\frac{dz}{L}\,
  \frac{\lvert f_m^{(\wp)}(z)\rvert^2}{p^2+k_m^2+\mathring{\tau}}
\end{equation}
for the two boundary conditions in question, 
$\wp=\text{per}$ and $\text{sp-sp}$, are given by
\begin{equation}\label{eq:I1per}
I^{(\mathrm{per})}_1(L;\mathring{\tau})=
\frac{A_{d-1}}{L}\,\mathring{\tau}^{(d-3)/2}-A_{d}\,
\mathring{\tau}^{(d-2)/2} 
 + \frac{2\,Q_{d,2}(\mathring{\tau}L^{2})}{\mathring{\tau}L^{d}}
\end{equation}
and
\begin{equation}\label{eq:I1sp}
I^{\mathrm{(sp-sp)}}_1(L;\mathring{\tau})
=I^{(\mathrm{per})}_1(2L;\mathring{\tau})\;,
\end{equation}
respectively. Here we started to use the notational conventions
summarized in Appendix~\ref{sec:conv} for momentum integrals such as
$\int_{\bm{p}}$. Further,
\begin{equation}\label{eq:Ad}
A_d=\frac{2\,N_d}{4-d}=-(4\pi)^{-d/2}\,\Gamma(1-d/2)\;,
\end{equation}
while
$Q_{d,2}(r)$ is a special one of the functions defined by\cite{rem:Qd2,CPT98}
\begin{equation}\label{eq:definitionQdsigma}
Q_{d,\sigma}(r)\equiv\frac{r}{2} \left[\sum_{k\in2\pi\mathbb{Z}}
  -\int_{-\infty}^\infty\frac{dk}{2\pi}\right]\int_{\bm{p}}\frac{(p^2+k^2)^{\sigma/2-1}}{p^2+k^2+r}\;.
\end{equation}

Using the definition~\eqref{eq:tbL}, the result implied by Eqs.~(3.20)--(3.24) of Ref.~\onlinecite{GD08} becomes
\begin{equation}
  \label{eq:gamma11om}
  \gamma^{(1;1)}(\bm{p},\omega)/\mathring{\lambda}= p^2-i\omega/\mathring{\lambda}+\mathring{\tau}^{(\wp)}_L(\mathring{\tau})+\mathcal{O}(\mathring{u}^2)\;.
\end{equation}

For a discussion of the properties of the functions $Q_{d,\sigma}$ the
reader is referred to Appendix D of Ref.~\onlinecite{GD08},
where also plots of $Q_{4,2}(r)$ and $Q_{6,2}(r)$ are displayed (see
Fig.~9 of this reference). It is known from previous
works\cite{KD92a,GD08} that the above results yield the
shifts~\eqref{eq:shifts} for $\mathring{\tau}=0$. This is easily verified
utilizing the fact that
\begin{eqnarray}
  \label{eq:limQd2ovr}
  \lim_{r\to0}Q_{d,2}(r)/r&=&
 \pi^{(d-5)/2}\, \Gamma[(3-d)/2]\, \zeta (3-d)/4\nonumber\\
&=&\pi^{-d/2}\,\Gamma(d/2-1)\,\zeta(d-2)/4
\end{eqnarray}
(when $d\ne 1,3,5$) according to equation (B35) of
Ref.~\onlinecite{DDG06} and (C9) or (A15) of
Ref.~\onlinecite{GD08}.

It is equally easy to verify that the contribution of graph (ii)
to $\gamma^{(1;3)}$ is in conformity with
Eq.~\eqref{eq:gamma13gamma4}. Upon expressing $R_L(t)$ in terms of
$\partial_t C_L$ via the fluctuation-dissipation
relation~\eqref{eq:FDT}, one arrives at the desired result $\int_0^\infty
dt\,R_L(t)\, C_L(t)=(2\mathring{\lambda})^{-1}[C_L(t=0)]^2$.

Let us also note that the RG equations~\eqref{eq:RGE} for the
functions $W^{(\tilde{N};N)}_R$ imply analogous ones for the
renormalized vertex functions $\gamma^{(\tilde{k};k)}_R$. In the case
of periodic boundary conditions, these vertex functions are
evidently multiplicatively renormalizable. One has 
$\gamma^{(\tilde{k};k)}_R
=Z_{\tilde{\phi}}^{\tilde{k}/2}Z_{\phi}^{k/2}\gamma^{(\tilde{k};k)}$,
which leads to the RG equations
\begin{equation}
  \label{eq:RGEgammas}
  \bigg[\mathcal{D}_\mu
-\tilde{k}\,\frac{\eta_{\tilde{\phi}}}{2}
-k\,\frac{\eta_\phi}{2}\bigg] 
\gamma^{(\tilde{k};k)}_R=0\,.
\end{equation}

For $\wp=\text{sp-sp}$ things are somewhat more subtle. It is known
that $\Gamma^{(1,3)}$ has uv singularities corresponding to
coun\-ter\-terms located on $\mathfrak{B}$ that are proportional to
$\tilde{\bm{\phi}}\cdot\partial_n\bm{\phi}$ and $
\bm{\phi}\cdot\partial_n\tilde{\bm{\phi}}$. Therefore, it is not
multiplicatively renormalizable and does not satisfy a homogeneous RG
equation.\cite{DD81b,Sym81,DD83a,Die86a}
However, when integrated with smooth background fields satisfying the
boundary conditions \eqref{eq:bc} with $\mathring{c}_1=\mathring{c}_2=0$, the boundary
singularities of the mentioned form give no contribution. The
construction of
$\mathcal{J}_{\text{eff}}[\tilde{\bm{\varphi}},\bm{\varphi}]$ involves
integrations with such functions (namely, $\psi\psi$-propagators).
Hence, the RG equations~\eqref{eq:RGEgammas}  carry over to the
case of sp-sp boundary conditions.

They can be exploited in a standard fashion to obtain the scaling
forms of the vertex functions
$\gamma^{(\tilde{k},k)}(\{\bm{p}_i,\omega_i\})$ for both types of
boundary conditions $\wp=\text{per}$ and $\wp=\text{sp-sp}$. One obtains
\begin{eqnarray}
  \label{eq:gammaRGsol}
  \lefteqn{\gamma_R^{(\tilde{k};k)}(\{\bm{p}_i,\omega_i\};\tau,L)\approx\lambda\, \mu^{d_\gamma-2}
  }&&\nonumber\\ &&\times
  (\mu\xi_\infty)^{-d_\gamma-\eta^*_\gamma}
  X^{(\wp)}_\gamma\big(\{\bm{p}_i\xi_\infty;\omega_i/\Omega_c\};L/\xi_\infty\big) \phantom{(4.23)}
\end{eqnarray}
with
\begin{eqnarray} 
   \label{eq:dgammaetc}
  d_{\gamma}&=&2-\tilde{k}+k-\frac{d-1}{2}(\tilde{k}+k-2)\;,\nonumber\\
\eta_\gamma^*&=&(\tilde{k}+k)\eta/2+(\tilde{k}-1)(\mathfrak{z}-2)/2\,,
\end{eqnarray}
and the characteristic  bulk frequency
\begin{equation}
\label{eq:Omegac}
  \Omega_c=\lambda\,(\mu\xi_\infty)^{-\mathfrak{z}}\,.
\end{equation}

\section{Scaling functions of finite-size susceptibilities} 
\label{sec:scffss}
We are now ready to turn to the computation of scaling functions by means
of the small-$\epsilon$ expansion. We first consider the scaling
function of the dynamical response function
$\mathring{\lambda}\langle\bm{\varphi}\cdot\tilde{\bm{\varphi}}\rangle$ at the bulk
critical point.
\subsection{Dynamical finite-size response function}
Noting that the $O(n)$ symmetry is unbroken when $T\ge T_{c,\infty}$,
we set $\alpha=\beta=1$ in the response and vertex functions
$W^{(1;1)}_{\alpha\beta}$ and $\gamma^{(1;1)}_{\alpha\beta}$ to obtain
\begin{equation}
  \frac{\mathring{\lambda}}{n}\,\langle\bm{\varphi}\cdot
  \tilde{\bm{\varphi}}\rangle=\frac{1}{L}\int_0^L dz\int_0^L dz'\,
  W^{(1;1)}_{11}\;.
\end{equation}
The $\bm{p}\omega$-transform of this quantity gives us the finite-size
susceptibility
\begin{eqnarray}
  \label{eq:chiLdef}
  \chi_L(p,\omega)&=&\int_0^L\frac{dz_1}{L}\int_0^Ldz_2
  \int d^{d-1}y_{12}\int dt_{12} \nonumber\\&&\strut \times \frac{\delta\langle\phi_1(\bm{x}_1,t_1)\rangle}{\delta
    h_1(\bm{x}_2,t_2)}\,e^{i(\omega t_{12}-\bm{p}\cdot\bm{y}_{12})}
\end{eqnarray}

The perturbation series of its inverse to order $o(\mathring{u}^2)$ can be
written as
\begin{eqnarray}
  \label{eq:chiL}
    [\chi^{(\wp)}_L(p,\omega)]^{-1} &=&\frac{-i\omega}{\mathring{\lambda}}
+p^2+\mathring{\tau}^{(\wp)}_L + \frac{n+2}{6}\,\frac{\mathring{u}}{L}\nonumber\\
&&\strut \times
G_\infty^{(d-1)}\big(\bm{0}|\mathring{\tau}_L^{(\wp)}\big) 
 +o(\mathring{u}^2)\;,\phantom{(5.3)}
\end{eqnarray}
where
\begin{equation}
  \label{eq:Gdmoneyy}
  G_\infty^{(d-1)}\big(\bm{0}|\mathring{\tau})=-A_{d-1}\,\mathring{\tau}^{(d-3)/2}
\end{equation}
is the static bulk propagator in $d-1$ dimensions
\begin{eqnarray}
  \label{eq:GGaussdmone}
  G_\infty^{(d-1)}(\bm{y}|\mathring{\tau})&=&\int_{\bm{p}}
    \,(p^2+\mathring{\tau})^{-1}\,e^{i\bm{p}\cdot\bm{y}} \nonumber\\ 
  &=&\frac{\left(\mathring{\tau}/y^2\right)^{(d-3)/4}}{(2\pi)^{(d-1)/2}}
  \,K_{(d-3)/2}\big(y\sqrt{\mathring{\tau}}\big)\,,\qquad
\end{eqnarray}
taken at zero separation $\bm{y}$.

From  the RG equations~\eqref{eq:RGE} one concludes that the
renormalized function $\chi_{L,R}^{(\wp)}$ has the asymptotic scaling
form 
\begin{equation}
  \label{eq:chiLsclform}
  \chi^{(\wp)}_{L,R}(p,\omega;\tau)\approx\mu^{-\eta}
  L^{2-\eta}\,\Xi^{(\wp)}(pL,\omega/\omega_L,L/\xi_\infty)\,, 
\end{equation}
where we have introduced the characteristic finite-size frequency
\begin{equation}
  \label{eq:omegaL}
  \omega_L=\lambda\,(\mu L)^{-\mathfrak{z}}\;.
\end{equation}

Using the above result~\eqref{eq:chiL}, one can verify that this
scaling form complies with the small-$\epsilon$ expansion. At the
order of our present calculation, the renormalization factors
$Z_\phi$, $Z_\lambda$, and $Z_u$ can all be substituted by unity,
but we need\cite{Die86a}
\begin{equation}
  \label{eq:Ztau}
  Z_\tau=1+\frac{n+2}{3\epsilon}\,u+O(u^2)
\end{equation}
to first order in $u$.  The pole term $\propto \mathring{\tau} u/\epsilon$ of
$\mathring{\tau}_L$ gets cancelled when $\mathring{\tau}$ and $\mathring{u}$ are expressed in terms of
renormalized quantities and the contribution from the counterterm
$\propto Z_\tau Z_\phi-1$ is added. The associated
renormalized shifts
\begin{equation}
  \label{eq:twpL}
  \tau^{(\wp)}_L=\mu^{-2}Z_\tau^{-1}\mathring{\tau}^{(\wp)}_L
\end{equation}
are uv finite; we obtain
\begin{eqnarray}
  \label{eq:tspL}
  \tau^{(\text{per})}_L&=&
  \tau^{(\text{sp-sp})}_{L/2}+O(u^2)\nonumber\\ &=&
\tau+\tau\,\frac{n+2}{6}\,\frac{
  u}{\tau^{\epsilon/2}}\bigg[\frac{\tau^{\epsilon/2}-1}{\epsilon/2}
+\frac{A_{d-1}}{\mu L\,\tau^{1/2}\,N_d} \nonumber\\ &&\strut
    + \frac{2\,Q_{d,2}(\mu^2\tau L^2)}{(\mu^2
  L^2\,\tau)^{d/2}\,N_d}  \bigg]+O(u^2)\;.
\end{eqnarray}

Combining the above results then gives
\begin{equation}
  \label{eq:chiLRres}
  [\chi_{L,R}^{(\wp)}(p,\omega)]^{-1}=\frac{-i\omega\mu^2}{\lambda}
+p^2+r_L^{(\wp)}+O(u^2)\;,
\end{equation}
where
\begin{eqnarray}
  \label{eq:rLperren}
  r^{(\wp)}_L
&=&\mu^2\tau^{(\wp)}_L\Big\{1-\frac{n+2}{6}\,\frac{u\,A_{3-\epsilon}}{\mu
  LN_{4-\epsilon}}\, \big[
\tau^{(\wp)}_L\big]^{-(1+\epsilon)/2}\Big\}\nonumber\\ &&\strut
+O(u^2) 
\end{eqnarray}
are the inverse static finite-size susceptibilities.

Note that by keeping the $O(u)$ terms of $\tau^{(\wp)}_L$ in the
contributions $\propto u\,[ \tau^{(\wp)}_L]^{(1-\epsilon)/2}$, we have
included contributions of order $u^2$ in these terms. While these
results for $r^{(\wp)}_L$ differ from those of Ref.~\onlinecite{GD08}
through precisely such terms, one easily checks that they  reduce to
Eqs.~(4.31) and (4.32) of this reference when expanded to first
order in $u$. 

Our rationale for keeping the $O(u)$ contribution to the shift in the
one-loop term associated with the zero mode should be clear. It
prevents the propagator $G^{(d-1)}_\infty$ that it involves from becoming
massless at bulk criticality when $L<\infty$. To compute the scaling
functions $\Xi^{(\wp)}$, we must compute $\chi_L^{(\wp)}$ at the
IR stable root $u^*=3\epsilon/(n+8)+O(\epsilon^2)$ of the beta
function $\beta_u$, where we can exploit the facts that both
$\mathfrak{z}-2$ and $\eta$ are of order $\epsilon^2$.  When $\tau>0$,
the shifted temperature variable $\mathring{\tau}^{(\wp)}_L$ is of zeroth order in $u$.
If we insert it into the contribution $\propto
u[\tau^{(\wp)}_L]^{(\epsilon-1)/2}$, the $O(\epsilon)$ term of
$\tau_L^{(\wp)}|_{u=u^*} $ will therefore produce an $O(\epsilon^2)$
contribution.  However, at criticality $\tau=0$, $\mathring{\tau}^{(\wp)}_L$ is linear in
$u$ and hence of order $u^*=O(\epsilon)$ when evaluated at $u^*$.
Accordingly, a contribution of order $\epsilon^{3/2}$ results from
this source.

Proceeding as indicated above, one arrives at the results
\begin{eqnarray}
  \label{eq:Chiscfcts}
  [\Xi^{(\wp)}(\mathsf{p},w,\mathsf{L})]^{-1}&=& \mathsf{p}^2
  -iw+X^{(\wp)}_\epsilon(\mathsf{L})\Big\{ 1\nonumber\\ &&\strut
  -2\pi\,\epsilon\,
  \frac{n+2}{n+8}\,\big[X^{(\wp)}_\epsilon(\mathsf{L})\big]^{-1/2}\Big\}   \qquad
\end{eqnarray}
with
\begin{eqnarray}
  \label{eq:XLeps}
  X_\epsilon^{(\text{per})}(\mathsf{L})&=&4\,X_\epsilon^{(\text{sp-sp})}(\mathsf{L}/2)
  \nonumber\\ &=&\mathsf{L}^2+\epsilon\,\frac{n+2}{n+8}\,\frac{16\pi^2\,Q_{4,2}(\mathsf{L}^2)+2\pi\,\mathsf{L}^3}{\mathsf{L}^2}\,.\phantom{(5.14)}
\end{eqnarray}

The latter functions have the asymptotic behaviors
\begin{equation}
  \label{eq:Xaslarge}
  X_\epsilon^{(\wp)}(\mathsf{L})\underset{\mathsf{L}\to\infty}{\approx}
    \mathsf{L}^2\big[1+O(\mathsf{L}^{-1})\big]
\end{equation}
and
\begin{equation}
  \label{eq:Xassmall}
X_\epsilon^{(\wp)}(\mathsf{L})\underset{\mathsf{L}\to 0}{\approx}
\frac{2\pi^2}{3a^2_\wp}\,\frac{n+2}{n+8}\,\epsilon\;,
\end{equation}
where $a_\wp$ is given by
\begin{equation}
  \label{eq:awp}
  a_\wp=
  \begin{cases}
    1&\text{for }\wp=\text{per}\,,\\
2&\text{for }\wp=\text{sp-sp}\,.
  \end{cases}
\end{equation}

Setting $\mathsf{p}=w=0$ in Eq.~\eqref{eq:Chiscfcts} gives us the scaling
functions of the static finite-size susceptibilities. One easily verifies
that our results for $\mathsf{L}^2/\Xi^{(\wp)}(0,0,\mathsf{L})$ are in
conformity with the $O(\epsilon)$ expressions~(4.46) and (4.47) of
Ref.~\onlinecite{GD08}. 

The limiting behavior~\eqref{eq:Xassmall} implies that the scaling
functions~\eqref{eq:Chiscfcts}  become 
\begin{eqnarray}
  \label{eq:dsuscscfct}
  [\Xi^{(\wp)}(\mathsf{p},w,0)]^{-1}&=&\mathsf{p}^2-iw
  +\frac{2\pi^2}{3\,a_\wp^2}\,\frac{n+2}{n+8}\bigg[\epsilon \nonumber\\
  &&\strut
  -\sqrt{6}\,a_\wp\,\Big(\frac{n+2}{n+8}\Big)^{1/2}\,\epsilon^{3/2}\bigg] 
\nonumber\\ &&\strut 
+o(\epsilon^{3/2})
\end{eqnarray}
at $\mathsf{L} =0$, which explicitly shows the announced $O(\epsilon^{3/2})$
contributions. 
Before we embark on a discussion of this result, let us note that the
scaling functions of the correlation function
\begin{eqnarray}
  \label{eq:CLdef}
  C^{(\wp)}_L(p,\omega)&=&\int_0^L\frac{dz_1}{nL}\int_0^Ldz_2\int d^{d-1}y_{12}\int
  dt_{12}\nonumber\\ &&\strut\times
  \langle\bm{\phi}(\bm{x}_1,t_1)\cdot\bm{\phi}(\bm{x}_2,t_2) 
\rangle^{\text{cum}}
  \,e^{i(\omega t_{12}-\bm{p}\cdot\bm{y}_{12})}\nonumber\\
\end{eqnarray}
need no separate calculations since the fluctuation-dissipation
relation~\eqref{eq:FDT} in conjunction with Eq.~\eqref{eq:chiLsclform}
yields
\begin{equation}
  \label{eq:Cscfkt}
  C^{(\wp)}_L(p,\omega)\approx \mu^{-\eta}
  L^{2-\eta}\,\frac{2}{w}\,\text{Im}\,\Xi^{(wp)}(\mathsf{p},w)\;.
\end{equation}

Had we computed the expansions~\eqref{eq:dsuscscfct} merely to linear
order in $\epsilon$, one might be tempted to think that even the
direct evaluation of these truncated series at $d=3$ would give
acceptable estimates. The $O(\epsilon^{3/2})$ terms reveal that there
is no reason for such optimism: their signs differ from those of the
contributions linear in $\epsilon$. Furthermore, their coefficients
are quite large. Thus, evaluating the series
expansion~\eqref{eq:dsuscscfct} to order $\epsilon^{3/2}$ at
$\epsilon=1$ yields (unphysical) \emph{negative} values even for $n=1$,
albeit the results remain positive when $\epsilon$ is sufficiently
small.  This shows that more sophisticated extrapolation techniques
are needed to obtain reliable estimates for these quantities at $d=3$.

An important check of the results for periodic boundary conditions can
be made by comparing them with the exact large-$n$ results obtained
from the solution of the mean spherical model with short range
interactions.\cite{BDT00,DDG06,CPT98} As we will show now, the
large-$n$ limit of the small-$\epsilon$
expansion~\eqref{eq:dsuscscfct} for $\wp=\text{per}$ can be recovered
by solving the self-consistent equation that the finite-size
susceptibility obeys in the limit $n\to\infty$ for $d=4-\epsilon$ in
an iterative fashion.

There are two reasons why a similar comparison with large-$n$ results
cannot be made here in the case of sp-sp boundary conditions. First,
translation invariance perpendicular to the boundary planes is broken
for these boundary conditions, just as it generally is for free
boundary conditions. The large-$n$ limit therefore corresponds to a
modified spherical lattice model involving separate
constraints\cite{Kno73,rem:msm} for the averages of $\sum_{i\in \text{layer
  }z}S_i^2 $ in each layer $z$, where $S_i$ is a spin variable on site
$i$. Second, when $d=3$, the thermal fluctuations should prevent the
occurrence of film and surface phases with long-range order at
temperatures $T>0$ in the continuous symmetry case $n>1$ and hence
for $n\to\infty$. Hence sp-sp boundary conditions cannot be realized
when $d=3$ and $n=\infty$. 

\subsection{Comparison with exact spherical-model results for periodic boundary conditions}

An important check of our results for periodic boundary conditions is their comparison with exact spherical-model results. 
The static finite-size susceptibility of the spherical model on a
$d$-dimensional slab with periodic boundary conditions can be written
as
\begin{equation}
  \label{eq:smsusc}
  \chi_L^{(\text{SM})}=L^2/X^{\text{SM}}_d(L/\xi_\infty)\;,
\end{equation}
where the scaling function $X^{\text{SM}}_d(\mathsf{L})$ is a solution
[see, eg., equations (5.2) and (4.66)  of Refs.~\onlinecite{GD08} and
\onlinecite{DDG06}, respectively]
\begin{equation}
\label{eq:fseqst}
  2\,Q_{d,2}(X^{\text{SM}}_d)/X^{\text{SM}}_d=A_d\,
  \big[\big(X^{\text{SM}}_d\big)^{(d-2)/2} -\mathsf{L}^{d-2}\big] \;.
\end{equation}

For $d=3$, the solution can be obtained in closed
form;\cite{DDG06,rem:DDG06} it reads
\begin{eqnarray}
  \label{eq:R0r3}
  X^{\text{SM}}_3(\mathsf{L}) &=& 4\,\mathrm{arccsch}^{2}
  \big( 2\,e^{-\mathsf{L}/2}\big) \nonumber\\ &=&
  4\ln^2\Big[\frac{1}{2}\Big(e^{\mathsf{L}/2}
  +\sqrt{4+e^{\mathsf{L}}}\Big) \Big]\,.
\end{eqnarray}
Its critical value, which corresponds to the amplitude
$1/\Xi^{\text{(per)}}(0,0,0)$, is\cite{CPT98}
\begin{equation}
  \label{eq:X0}
  X^{\text{SM}}_3(0)=4\ln^2g\simeq 0.926259\;,\quad
  g=\big(1+\sqrt{5}\,\big)/2\;,
\end{equation}
where $g$ is the golden mean.

To determine the amplitude $X^{\text{SM}}_{d}(0)$ in $d=4-\epsilon$
dimensions, we substitute the series expansion of the function
$Q_{d,2}(r)$ given in equation (C.9) of Ref.~\onlinecite{GD08} to
obtain the representation
\begin{equation}
  \label{eq:2Qd2Aexp}
  \frac{2\,Q_{4-\epsilon,2}(r)}{r}=
  A_{4-\epsilon}\,r^{1-\epsilon/2}-A_{3-\epsilon}\,r^{(1-\epsilon)/2}
  +f_\epsilon(r)  
  \end{equation}
with
\begin{equation}
\label{eq:feps}
f_\epsilon(r)=\sum_{j=0}^\infty
b^{(j)}_\epsilon\,r^j \;,
\end{equation}
where
\begin{equation}
\label{eq:beps}
  b^{(j)}_\epsilon=\frac{(-1)^j\, \Gamma[j+(\epsilon-1)/2]}{j!\, 2^{2
      j+1}\, \pi^{2j+(\epsilon +1)/2}}\,\zeta(2j+\epsilon-1)\;.
\end{equation}

The coefficients $A_{4-\epsilon}$ and $b_\epsilon^{(1)}$ have simple
poles at $\epsilon=0$ whose residues differ by the factor $(-2)$, so that
the pole term of $A_{4-\epsilon}\,r^{1-\epsilon/2}$ cancels that of
$b^{(1)}_\epsilon r$. The remaining coefficients, $A_{3-\epsilon}$ and
$b_\epsilon^{(j)}$ with $j\ne 1$, are regular at $\epsilon=0$. On the
other hand, $A_{3-\epsilon}$ and $b^{(0)}_\epsilon$ have simple poles at
$\epsilon=1$ whose residues are equal.

These results imply that $X_{4-\epsilon}(\mathsf{L})-\mathsf{L}^2$
must vanish linearly as $\epsilon\to 0$, so that the right-hand side
of Eq.~\eqref{eq:fseqst} approaches a nonzero limit. The $O(\epsilon)$
term of $X_{4-\epsilon}(\mathsf{L})-\mathsf{L}^2$ was computed in
Ref.~\onlinecite{GD08} [see its Eq.~(5.3)]. To determine the
$O(\epsilon^{3/2})$ contribution to $X_{4-\epsilon}(0)$, we set
$\mathsf{L}=0$ and solve Eq.~\eqref{eq:fseqst} with the ansatz
$X_{4-\epsilon}(0) = C_1\epsilon+ C_{3/2}\epsilon^{3/2}
+o(\epsilon^{3/2})$. This gives
\begin{equation}
  X^{\text{SM}}_{4-\epsilon}(0)=\frac{2\pi^2}{3}\big(\epsilon - \sqrt{6}\,
  \epsilon^{3/2}\big)+o(\epsilon^{3/2}) \;.
\end{equation}
Obviously, the result agrees with the expansion~\eqref{eq:dsuscscfct}
of $[\Xi^{(\text{per})}(\mathsf{p},w,0)]^{-1}$ for $n=\infty$.

\subsection{Dynamical finite-size surface response function}
\label{sec:scfs} 

In Sec.~V~B we were concerned with the dynamic
finite-size susceptibilities~\eqref{eq:chiLdef}. These are integral
quantities. As an alternative, we here wish to consider the local
susceptibilities $\chi^{(\text{sp-sp})}_{11} = \mathring{\lambda}\,\langle
\bm{\phi}^{\mathfrak{B}_1}(\bm{y},t)\cdot
\tilde{\bm{\phi}}^{\mathfrak{B}_1}(\tilde{\bm{y}},\tilde{t\,})\rangle/n$
and $\chi^{(\text{sp-sp})}_{1L} = \mathring{\lambda}\,\langle
\bm{\phi}^{\mathfrak{B}_1}(\bm{y},t)\cdot
\tilde{\bm{\phi}}^{\mathfrak{B}_2}(\tilde{\bm{y}},\tilde{t\,})\rangle/n$
describing the linear responses of the order-parameter density at one
boundary plane to a magnetic field acting on the order-parameter
density in the same and the complementary boundary planes,
respectively. In the case of periodic boundary conditions, both
quantities are the same --- and by translational invariance along the
$z$-direction --- identical with the layer susceptibility
$\chi_{zz}=\mathring{\lambda}\,\langle \bm{\phi}(\bm{y},z,t)\cdot
\tilde{\bm{\phi}}(\tilde{\bm{y}},z,\tilde{t\,})\rangle/n$ for any
layer $z$. We start with an investigation of the latter. For the sake of
simplicity, we do not consider deviations from the bulk critical
temperature here, setting $\tau=0$.

\subsubsection{Layer susceptibility for periodic boundary conditions}
\label{sec:lspbc}

Our analysis of the finite-size susceptibility $\chi_L$ suggests that
we should work with a dressed $\varphi\varphi$-propagator that
accounts for the $O(\mathring{u})$ shift $\mathring{\tau}^{(\text{per})}_L$ given in
Eq.~\eqref{eq:shifts}. Noting that the $k_0=0$-mode contribution to
$C_L(\bm{p}=\bm{0};z=0;t=0)$ vanishes when $\mathring{\tau}=0$, one sees that a
single two-loop graph suffices to capture all contributions to order
$\epsilon^{3/2}$ in the renormalized quantity $\chi_{zz,R}$. One obtains
\begin{eqnarray}
  \label{eq:chizzpt}
\chi_{zz}(p,\omega;L)/\mathring{\lambda}&=&\frac{\coth(\mathring{\kappa}_\omega
  L/2)}{2\mathring{\lambda} \mathring{\kappa}_\omega}
+\raisebox{-3mm}{\includegraphics[scale=0.8]{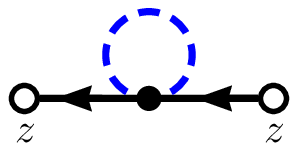}}\nonumber\\
&&\strut +
\raisebox{-3mm}{\includegraphics[scale=0.8]{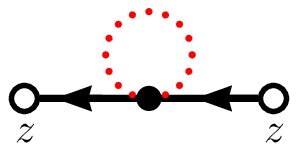}} +O(\mathring{u}^2)\;,
\end{eqnarray}
where $\mathring{\kappa}_\omega$ now denotes the
quantity~\eqref{eq:kapom} with $\mathring{\tau}$ set to zero. Here full black
lines with and without arrows represent the free response and
correlation propagators $R_L$ and $C_L$ with $\mathring{\tau}=0$, respectively.
The dashed blue line indicates $C^{(\text{per})}_{L,\psi}$ with $\mathring{\tau}=0$.
The dotted red line is the $k=0$ part of $C^{(\text{per})}_L$ with
$\mathring{\tau}=\delta\mathring{\tau}^{(\text{per})}_L$. Note that we do not
include the shift $\delta\tau^{(\text{per})}_L$ in the
$k=0$~contributions to the external legs of the two graphs. Their
inclusion would produce terms of order $\mathring{u}^2$ and hence lead to
corrections of order $\epsilon^2$.

In Appendix~\ref{sec:graphs} the graphs in Eq.~\eqref{eq:chizzpt} are
computed. From the results given in Eqs.~\eqref{eq:oneloopchizzpsi}
and \eqref{eq:oneloopchizzvarphi}, the RG-improved perturbation
expansion of the renormalized dynamic layer susceptibility
$\chi_{zz,R}=\chi_{zz}/ Z_\phi$ follows in a straightforward
fashion. Evaluating it at the fixed-point value, one sees that it is
in conformity with the scaling form predicted by the RG
equations~\eqref{eq:RGE}, namely
\begin{equation}
  \label{eq:chizzscf}
  \chi^{(\text{per})}_{zz,R}(p,\omega;L,\tau=0)\approx \mu^\eta
  L^{1-\eta} \,\Xi_{zz}^{(\text{per})}(pL,\omega/\omega_L)\;.
\end{equation}
where the scaling function has the  small-$\epsilon$ expansion
\begin{eqnarray}
  \label{eq:Xizzper}
  \Xi_{zz}^{(\text{per})}(\mathsf{p},w)&=&
  \frac{\coth(\varkappa/2)}{2\varkappa}
  -\frac{\pi^2}{12}\, \frac{n+2}{n+8}
  \,\frac{\varkappa+\sinh\varkappa}{\varkappa^3\sinh^2(\varkappa/2)}
  \nonumber\\ && \times \left[\epsilon-
    \sqrt{6\,\frac{n+2}{n+8}}\,\epsilon^{3/2}\right]
  +o(\epsilon^{3/2})\phantom{(5.23)}  
\end{eqnarray}
with
\begin{equation}
  \label{eq:varkappa}
  \varkappa=\sqrt{\mathsf{p}^2-iw}\;.
\end{equation}

The susceptibility $\chi_{zz,R}$ must reduce to an $L$-independent
bulk quantity as $L\to \infty$. Therefore, the scaling 
function should have the limiting behavior
\begin{equation}
  \label{eq:scflargep}
   \Xi_{zz}^{(\text{per})}(\mathsf{p},w)
   \underset{\mathsf{p}\to\infty}{\approx}\mathsf{p}^{\eta-1}\, \Psi(w
   \mathsf{p}^{-\mathfrak{z}})  \;.
\end{equation}
Equation~\eqref{eq:scflargep} shows that this is indeed the case, giving
\begin{equation}
  \label{eq:Psi}
  \Psi(w\mathsf{p}^{-\mathfrak{z}})
  =\frac{1}{2\sqrt{1-iw\mathsf{p}^{-\mathfrak{z}}}}+O(\epsilon^2) \;,
\end{equation}
which is the correct bulk result.

As long as $L$ is finite, there is no reason for $\chi_{zz,R}$ to
diverge at the bulk critical temperature. Hence, we expect its inverse
$1/\chi_{zz,R}(p,0;L,0)$ to approach a nonzero limit. From
Eq.~\eqref{eq:Xizzper} we can read off  the small-$\epsilon$  expansion of 
$1/\Xi_{zz}^{(\text{per})}(\mathsf{p},w)$ and compute its limit
\begin{eqnarray}
  \label{eq:scfsmallp}
\lim_{\mathsf{p}\to 0}
\frac{1}{\Xi_{zz}^{(\text{per})}(\mathsf{p},0)}&=&
\frac{2\pi^2}{3}\,\frac{n+2}{n+8}\left[\epsilon-
  \sqrt{6\,\frac{n+2}{n+8}}\,\epsilon^{3/2}\right] \nonumber\\ &&\strut 
+o(\epsilon^{3/2})\,.
\end{eqnarray}
The result agrees at this order with the
expansion~\eqref{eq:dsuscscfct} of the inverse static susceptibility
$[\Xi^{(\text{per})}(0,0,0)]^{-1}$. Thus the inverse static 
zero-momentum layer susceptibility $1/\chi_{zz,R}(0,0;L,0)$ is indeed
nonzero. Of course, the previously discussed difficulties to
obtain reliable $d=3$ estimates from this expansion to the order
$\epsilon^{3/2}$ apply here as well. 

\subsubsection{Surface susceptibilities for sp-sp boundary conditions}
\label{sec:sscspsp}

The graphs of the surface susceptibilities
$\chi^{(\text{sp-sp})}_{11}$ and $\chi^{(\text{sp-sp})}_{1L}$
corresponding to the ones displayed in Eq.~\eqref{eq:chizzpt} are
computed in Appendix~\ref{sec:spspbc}. The results are gathered in
Eqs.~\eqref{eq:oneloopchi11psi}--\eqref{eq:oneloopchi11varphi}. Using
them one can verify in a straightforward manner that the renormalized
susceptibilities
$\chi^{(\text{sp-sp})}_{11,R}=\chi^{(\text{sp-sp})}_{11}/Z_\phi Z_1$
and $\chi^{(\text{sp-sp})}_{1L,R}=\chi^{(\text{sp-sp})}_{1L}/Z_\phi
Z_1$ are uv finite to the appropriate order in $u$. One obtains
\begin{eqnarray}
  \label{eq:chi11Rres}
  \chi^{(\text{sp-sp})}_{11,R}&=&\frac{\coth(\kappa_\omega
    L)}{\kappa_\omega}-\frac{n+2}{6}\,u\bigg\{\frac{\coth(\kappa_\omega L)}{\kappa_\omega}\,\Big[ 1-2\gamma_E\nonumber \\ &&\strut -2\ln (\mu L)]  
+S(\kappa_\omega L)
+R(\kappa_\omega L)\nonumber\\ &&\strut 
 +\frac{\pi^2}{3}\bigg[1-\frac{4}{L}\,\bigg(\frac{n+2}{2\mu^2L^2}\,
   u\bigg)^{1/2}\bigg]
I_{1,1}(\kappa_\omega L)
\nonumber\\ &&\strut
 +O(\epsilon)\bigg\}
\end{eqnarray}
and
\begin{eqnarray}
  \label{eq:chi1LRres}
  \chi^{(\text{sp-sp})}_{1L,R}&=&\frac{\text{csch}(\kappa_\omega
    L)}{\kappa_\omega}-\frac{n+2}{6}\,u\bigg\{\frac{\text{csch}(\kappa_\omega
  L)}{\kappa_\omega}\,[ 1-2\gamma_E
 \nonumber \\ &&\strut 
 -2\ln (\mu L)]  
 +2S_{1,2}^{(0)}(\kappa_\omega L) +R_2(\kappa_\omega L)
\nonumber\\ &&\strut
 +\frac{\pi^2}{3}\bigg[1-\frac{4}{L}\,\bigg(\frac{n+2}{2\mu^2L^2}\,
   u\bigg)^{1/2}\bigg]I_{1,2}(\kappa_\omega L) 
\nonumber\\ &&\strut +O(\epsilon)\bigg\}
 \,,
\end{eqnarray}
where $\kappa_\omega=\sqrt{p^2 -i\omega\mu^2/\lambda}$, while
$S(\varkappa)$, $R(\varkappa)$, $I_{1,1}(\varkappa)$,
$S_{1,2}^{(0)}(\varkappa)$, $R_{2}(\varkappa)$, and
$I_{1,2}(\varkappa)$ denote the functions specified by
Eqs.~\eqref{eq:Sdef}, \eqref{eq:Rdef}, \eqref{eq:Ijjres},
\eqref{eq:S120},  \eqref{eq:R2}, and
\eqref{eq:I12res}, respectively. The functions $R(\varkappa)$ and
$R_2(\varkappa)$ are single integrals, which can be computed by
numerical integration.

According to the RG equations~\eqref{eq:RGE} these susceptibilities
should have the asymptotic scaling behavior
\begin{equation}
  \label{eq:chi1jscbeh}
  \chi^{(\text{sp-sp})}_{\substack{11,R\\ 1L,R}}(p,\omega;L,\tau=0)
\approx\mu^{-\eta_{\|}^{\text{sp}}}L^{1-\eta_{\|}^{\text{sp}}}
\,X_{\substack{11\\ 1L}}(pL,\omega/\omega_L)\;,
\end{equation}
where $\eta_{\|}^{\text{sp}}$ is a standard surface correlation exponent
associated with the special transition whose $\epsilon$~expansion
$\eta_{\|}^{\text{sp}}=-\epsilon\,(n+2)/(n+8)+O(\epsilon^2)$ is
known\cite{DD81b,DD83a,Die86a} to order $\epsilon^2$.
Evaluating  the above results~\eqref{eq:chi11Rres} and
\eqref{eq:chi1LRres} at $u^*$ shows their consistency with these
scaling forms and yields the small-$\epsilon$ expansions
\begin{eqnarray}
  \label{eq:X11}
  X_{11}(\mathsf{p},w)&=&\frac{\coth\varkappa}{\varkappa}-\frac{n+2}{n+8}\,\frac{\epsilon}{2}\,\bigg\{(1-2\gamma_E)\,\frac{\coth\varkappa}{\varkappa}\nonumber\\ && \strut +S(\varkappa)+R(\varkappa)
\nonumber\\ &&\strut +\frac{\pi^2}{3}\,
 \bigg[1-2\Big(6\epsilon\,\frac{n+2}{n+8}\Big)^{1/2}\bigg]I_{1,1}(\varkappa)\nonumber\\ &&\strut +o(\epsilon^{3/2}) \bigg\}
\phantom{(5.31)}
\end{eqnarray}
and
\begin{eqnarray}
  \label{eq:X1L}
  X_{1L}(\mathsf{p},w)&=&\frac{\text{csch}\varkappa}{\varkappa}-\frac{n+2}{n+8}\,\frac{\epsilon}{2}\,\bigg\{(1-2\gamma_E)\,\frac{\text{csch}\varkappa}{\varkappa}\nonumber\\ && \strut +2S^{(0)}_{1,2}(\varkappa)+R_2(\varkappa)
\nonumber\\ &&\strut +\frac{\pi^2}{3}\,
 \bigg[1-2\Big(6\epsilon\,\frac{n+2}{n+8}\Big)^{1/2}\bigg]I_{1,2}(\varkappa)\nonumber\\ &&\strut +o(\epsilon^{3/2}) \bigg\}
\end{eqnarray}
where $\varkappa$ is again given by Eq.~\eqref{eq:varkappa}.

In the limit $\mathsf{p}\to\infty$, the function $X_{11}$ must behave as
\begin{equation}
  \label{eq:X11largep}
  X_{11}(\mathsf{p},w)\underset{\mathsf{p}\to\infty}{\approx}\mathsf{p}^{\eta_\|^{\text{sp}}-1}\,\Psi_{11}(w\mathsf{p}^{-\mathfrak{z}})\,,
\end{equation}
where $\Psi_{11}$ must agree with the scaling function of
$\chi^{(\text{sp-sp})}_{11,R}$ of the semi-infinite ($L=\infty$)
theory. This is indeed the case. The limit
$\lim_{\mathsf{p}\to\infty}\mathsf{p}^{1-\eta_{\|}^{\text{sp}}}
\Psi_{11}(\mathsf{p},w)$ 
can be computed in a straightforward fashion, giving
\begin{equation}
  \label{eq:Psi11}
  \Psi_{11}(v)=\frac{1}{\sqrt{1-iv}}\Big\{1
  +\frac{\epsilon}{2}\,\frac{n+2}{n+8}
  \,[1-\ln(4-4iv)]\Big\}+O(\epsilon^2)\;.  
\end{equation}
An independent calculation for the semi-infinite theory yields
precisely the same result.

Since $\chi^{(\text{sp-sp})}_{1L,R}$ vanishes as $L\to\infty$, the
analog of the scaling function $\psi_{11}$ for
$\chi^{(\text{sp-sp})}_{1L,R}$ must vanish. Checking the limit
$\lim_{\mathsf{p}\to\infty}\mathsf{p}^{1-\eta_{\|}^{\text{sp}}}
\Psi_{1L}(\mathsf{p},w)$ explicitly, we do in fact find that it is
zero. 
 
Just as the layer susceptibility $\chi_{zz,R}$, the surface
susceptibilities $\chi^{(\text{sp-sp})}_{11,R}$ and
$\chi^{(\text{sp-sp})}_{1L,R}$ should not become critical at the bulk
critical point and hence have finite $p\to 0$ limits at $\tau=\omega$.
To check this, we set $\omega=0$ in Eqs.~\eqref{eq:X11} and
\eqref{eq:X1L}, determine the series expansions of
$[X_{11}(\mathsf{p},0)]^{-1}$ and $[X_{1L}(\mathsf{p},0)]^{-1}$ to
$O(\epsilon^{3/2})$, and then take the limits $\mathsf{p}\to0$. The
results are
\begin{eqnarray}
  \label{eq:X11Rinv0}
  \lefteqn{[X_{11,R}(0,0)]^{-1}= [X_{1L,R}(0,0)]^{-1}+o(\epsilon^{3/2})}&&\nonumber\\
&=&\epsilon\,
  \frac{\pi^2}{6}\,\frac{n+2}{n+8}
  \left[1-2\Big(6\epsilon\,\frac{n+2}{n+8}\Big)^{1/2}\right]
  +o(\epsilon^{3/2})\,. \phantom{(5.33)}
\end{eqnarray}
They indicate that an $L$-dependent shift of the multicritical special
point occurs also along the $c$-direction, not only along the
temperature direction. Note, however, that a special point at which a
bulk disordered phase with long-range surface order meets with a
bulk-disordered, surface-ordered phase and a bulk-ordered,
surface-ordered phase exists for $d=3$ dimensional, semi-infinite spin
systems only when $n=1$. The analog of this multicritical point for
finite $L$ is the one where the transition temperature of the surface
transition coincides with the transition temperature $T_{c,L}$ of the
film.

\section{Conclusions}
\label{sec:concl}

In this article, we have studied model $A$ in film geometry. We focused
on the cases of boundary conditions for which the classical (Landau-van
Hove) theory involves zero modes: periodic boundary conditions and free
boundary conditions with critically enhanced surface interactions.

Major motivations were to determine the consequences of the zero mode and
to reformulate RG-improved perturbation theory such that a
well-defined small-$\epsilon$ expansion results at the bulk critical
temperature. Our reformulation of RG-improved perturbation theory
builds on the strategy pursued in Refs.~\onlinecite{DGS06} and
Ref.~\onlinecite{GD08} to investigate static properties such as the
finite-size free energy and the Casimir force. In these papers it was
shown that the conventional $\epsilon$-expansion breaks down and 
contributions involving half-integer powers $\epsilon^{k/2}$ with
$k\ge 3$ modulo powers of $\ln \epsilon$ appear in the expansion of
the critical Casimir amplitudes and other quantities. Here, we
confirmed this breakdown and explicitly determined the
small-$\epsilon$ expansions of various dynamic finite-size, layer, and
surface susceptibilities at $T_{c,\infty}$ to order $\epsilon^{3/2}$.
We were also able to cast the $\epsilon$-expansion of the
temperature-dependent scaling functions of the finite-size
susceptibilities $\chi^{(\wp)}_{L,R}$ for both boundary conditions
$\wp=\text{per}$ and $\wp=\text{sp-sp}$ in a form that reproduce their
expansions to order $\epsilon^{3/2}$ as $T\to T_{c,\infty}$.

Despite these successes, the theory suffers from severe limitations
and is still in an unsatisfactory state. First of all, our results
\eqref{eq:scfsmallp} and \eqref{eq:X11Rinv0} for the inverse layer and
surface susceptibilities at $T_{c,\infty}$ suggest that reliable
estimates for $d=3$ can hardly been obtained on the basis of such
small-$\epsilon$ expansion to order $\epsilon^{3/2}$ alone. Just as in
the case of the Casimir amplitudes and the scaling function of the
residual free energy at small values of $L/\xi_\infty$,\cite{GD08} the
deviations of the simplest extrapolations obtained by setting
$\epsilon=1$ in the series truncated at the respective orders
$\epsilon$ and $\epsilon^{3/2}$ appear to oscillate in sign. It is
conceivable that the situation is particularly bad in those cases
where the classical theory involves zero modes. These zero modes
provide a separate source of IR singularities, a problem one also
encounters in thermal field theory.\cite{LeB04} It has been suggested
to handle IR problems of this kind by a resummation of ``foam''
diagrams.  In the case of the $n$-component bulk $\phi^4$ theory at
finite temperature, \cite{DHLR98} this procedure yields results for
the pressure in conformity with the exact large-$n$ result. Whether
and to what extent RG-improved perturbation theory might be improved
by combining it with resummations of this kind remains to be seen. 

Finally, let us emphasize that there is little reason to believe that
the theory is in a much better state in those cases where the
classical theory does not involve zero modes, notwithstanding the
additional difficulties such modes cause. For free boundary
conditions, one will generically encounter a zero mode in the
classical theory at an $L$-dependent temperature $T^{(0)}_{c,L}\ne
T_{c,\infty}$. This indicates that in Landau theory the film becomes
critical at this temperature and undergoes a transition to an ordered
low-temperature phase. The ordered phase and hence the transition to
it may not survive the inclusion of thermal fluctuations when
$L<\infty$. This happens in the continuous symmetry case $n>1$ when
$d\le 3$, where it should be recalled that the $d=3$ case with $n=2$
is special in that the film has a low-temperature phase with
quasi-long-range order. However, even in those cases where the film
does have a transition to an ordered low-temperature phase for finite
$L$ (as it does when $d=3$ and $n=1$), one encounters two important
challenges that are beyond the scope of the presently available
analytical RG approaches but any satisfactory full theory of
dimensional crossover must be able to cope with: (i) to determine the
location of the singularity of the residual free energy's scaling
function corresponding to the transition temperature $T_{c,L}$ with
acceptable accuracy, and (ii) to yield the correct IR
singularities at this transition in conformity with the expected
$(d-1)$-dimensional critical behavior of the film. The difficulty is
that even the shift cannot normally be computed by perturbation theory
but requires RG techniques to deal with the IR
singularities.\cite{rem:shift} The RG scheme employed here and in the work of
Krech and Dietrich\cite{KD92a} may be appropriate to go on
scales of the order of $\xi_\infty$ as long as $\xi_\infty\lesssim L$.
However, it is insufficient to integrate out degrees of freedom
between $L$ and the film's correlation length $\xi_L$ in an adequate
fashion when $\xi_L>L$, and to correctly yield the IR singularities at
the critical temperature $T_{c,L}$ even when the boundary conditions
do not involve zero modes at $T_{c,\infty}$.

We close with some comments on the universality of our results and
finite-size scaling results in general. Chen and
Dohm\cite{CD02b,Doh08} recently launched a discussion of the
universality of finite-size scaling results and the validity of
two-scale (and multiscale) factor universality. 
Let us consider their concerns in some detail.  A first issue raised in
Ref.~\onlinecite{CD02b} is that the use of a sharp large-momentum
cutoff modifies the $L$~dependence of the singular part of the
finite-size free energy density of systems of linear size $L$ in a
qualitative manner. This effect is unphysical and entirely due to the
use of a sharp cutoff; that a sharp cutoff can produce unphysical
effects has been known since the early days of Wilson's RG.\cite{WK74}
For systems of the kind considered by them --- systems that are finite
in all directions --- the issue was discussed and clarified in
Refs.~\onlinecite{DR01} and \onlinecite{DKD03}. It needs no further
discussion.

A second point made in Refs.~\onlinecite{CD02b} is that long-range
interactions which are irrelevant in the RG sense produce algebraically
decaying contributions to the singular part of the finite-size free
energy density. Such interactions were previously considered by
Dantchev and co-workers,\cite{DR01,chamati2002b} who introduced the term ``subleading
long-range interaction'' for them. Away from criticality, these
contributions compete with the exponentially decaying ones one has for
systems with purely short-range interactions and become dominant in
the appropriate region of temperature and large $L$. While such
contributions (which are expected to be small on an absolute scale)
still have to be clearly verified by experiments, they certainly are
real.

Chen and Dohm\cite{CD02b} interpreted their presence as signaling the
breakdown of finite-size scaling. However, what is broken is just the
simple version of finite-size scaling that involves a \emph{single
  length} besides $L$, namely, the correlation length. Subleading
long-range interactions give rise to at least one further length ---
the one associated with the corresponding irrelevant scaling
field.\cite{rem:subllri} This \emph{must not be set to zero} in order
to retain the long-range tail in the regime where it dominates the
exponentially decaying short-range contribution to the finite-size
free-energy density. It may well be set to zero in the critical regime
$L/\xi_\infty\ll 1$ because the subleading long-range interaction
contributes there only a correction to the leading $L$ dependence.
Thus, subleading long-range interactions are intermediate between
dangerous irrelevant and conventional irrelevant perturbations: they
share with the former the property that they must not generally be set
to zero.  Unlike those (which would affect the leading critical
behavior of some quantities), but similar to the latter, they give
only corrections to the leading critical behavior.

Note that the mechanism just described for subleading long-range
interactions is neither specific to finite-size critical behavior nor
new. A familiar analog known from the study of critical adsorption of
fluids was discussed more than 25 years ago by de Gennes.\cite{deG81}
Substrates (``walls'') typically exert one-body forces on the fluid
that have besides short-range components algebraically decaying
van-der-Waals tails. In a semi-infinite geometry bounded by a wall at
$z=0$ and restricted to $z\ge 0$, the latter contribute effective
wall-fluid interactions of the form $\int d^{d-1}y\int_0^\infty
dz\,h_{\text{wf}}(z)\,\phi(\bm{y},z)$ to the Hamiltonian, where
$h_{\text{wf}}(z)$ behaves as $ A_{\text{wf}}\,z^{-\upsilon}$ as
$z\to\infty$ (cf.\ Ref.~\onlinecite{PL83} and Sec.~3.11 of
Ref.~\onlinecite{Die86a}). The long-range part is irrelevant in the RG
sense provided the exponent $\upsilon$ is larger than the magnetic RG
eigenexponent $y_h=(d+2-\eta)/\nu$, which is the case for nonretarded
and retarded van-der-Waals interactions in $d=3$ dimensions (for which
$\upsilon=3$ and $\upsilon=4$, respectively). It produces a long-range
tail $\sim z^{-\upsilon}$ to the deviation $\delta
m(z)=m(z)-m(\infty)$ of the order-parameter density
$m(z)=\langle\phi(\bm{y},z)\rangle$ from its bulk value $m(\infty)$.
By setting the length $\sim g_{\text{wf}}^{\upsilon-y_h}$ associated
with the irrelevant scaling field $g_{\text{wf}}\sim A_{\text{wf}}$ to
zero, one would loose this algebraically decaying contribution. On the
other hand, the leading temperature singularity $\sim
|\tau|^{-(\nu-\beta)}$ of the excess order parameter $\int_0^\infty
dz\,\delta m(z)$ would remain the same since $g_{\text{wf}}$ yields
merely corrections to scaling for this quantity. The analogy with how
subleading long-range interactions affect finite-size properties is
obvious.  Of course, the amplitude of the irrelevant scaling field
$g_{\text{wf}}$ is nonuniversal. Following the logic of
Refs.~\onlinecite{CD02b} and \onlinecite{Doh08}, one would have to
call this a violation of scaling in semi-infinite systems, though it again just
means that single-length scaling reaches its limits,
failing to capture the asymptotic behavior of certain quantities.

To what extent would the inclusion of subleading long-range
interactions alter the results of our analysis of dynamic finite-size
critical behavior given above? While a detailed, quantitative analysis
of their effects is beyond the scope of this paper, clear predictions
can be made on general grounds and the basis of what is known from
statics. Power laws describing asymptotic dependences in $L$ or $\tau$
at $T=T_{c,\infty}$ will be modified by corrections to scaling
involving the associated irrelevant scaling field.\cite{rem:subllri}
For a subleading long-range pair interaction decaying $\sim
x^{-(d+\sigma)}$ (with $\sigma>2-\eta$), the associated
correction-to-scaling exponent is $\omega_\sigma=\sigma-2+\eta$ (see,
e.g., Ref.~\onlinecite{DDG06} and its references). Thus these
corrections should be down by factors $L^{-\omega_\sigma}$ (or
$\xi_\infty^{-\omega_\sigma}$) in comparison to the respective leading
power laws. Correspondingly, scaling forms such as the one for the
finite-size susceptibility $\chi^{(\wp)}_{L,R}(p,\omega;\tau)$ given
in Eq.~\eqref{eq:chiLsclform} should obtain a correction $\propto
L^{2-\eta-\omega_\sigma}\,
\Xi_\sigma^{(\wp)}(pL,\omega/\omega_L,L/\xi_\infty)$ at linear order
in the irrelevant long-range scaling field.\cite{rem:subllri}
Furthermore, there exist quantities whose behaviors get
\emph{qualitatively modified} by subleading long-range interactions.
This is typically the case (in certain regimes of $L$ and $\tau$) for
quantities that decay exponentially in the absence of long-range
interactions. Obvious examples are zero-frequency response functions
in position space at temperatures $\tau>0$; these decay algebraically
in the limit of large distances $x_{ij}=|\bm{x}_i-\bm{x}_j|\to
\infty$ between two points when the pair interactions have a
subleading long-range tail.

Of course, a proper investigation of the effects of subleading
long-range interactions should also allow for irrelevant
surface-related scaling fields (such as pair interactions localized on
the boundary that decay algebraically as a function of the separation
$\bm{y}_i-\bm{y}_j$ along the boundary planes and pair interactions in
the interior of the sample that decay as a power of the distance from
the boundary planes). This is beyond the scope of our present work. 
                      
We conclude by turning to a third source of universality violations,
discussed extensively in Ref.~\onlinecite{Doh08}: the effects of weak
anisotropy. To keep things as simple as possible, it will be
convenient to discuss the issue first in the context of static bulk
critical behavior. The characteristic property of systems exhibiting
weakly anisotropic \emph{bulk} critical behavior is that the
correlation lengths describing the decay of correlations along
arbitrary directions diverge $\sim |\tau|^{-\nu}$ with one and the
same critical exponent $\nu$, but the shape of the correlation region
is ellipsoidal rather than spherical.  This means that the square
gradient term of the Hamiltonian~\eqref{eq:Ham} in general takes the
form
\begin{equation} \label{eq:modsgt}
  \frac{1}{2}\int_{\mathfrak{V}}d^dx\,B^{kl}\,(\partial_k\bm{\phi})
  \cdot\partial_l\bm{\phi}  
\end{equation}
in Cartesian coordinates, where
$\partial_k\bm{\phi}=\partial\bm{\phi}/\partial x^k$ are partial
derivatives with respect to these Cartesian coordinates
$x^k,\,k=1,\dotsc,d$, and Einstein's summation convention is used. The
matrix $\bm{B}\equiv (B^{kl})$ is symmetric and positive definite. Hence its inverse exists
and defines a metric tensor $B_{kl}$. Accordingly the modified square gradient
term~\eqref{eq:modsgt} can be viewed as the scalar product of the gradient operator with itself in this metric.

An evident consequence of the choice~\eqref{eq:modsgt} of the
modified square gradient term is that a corresponding replacement 
\begin{equation}
  \label{eq:repnablanabla}
  \overleftarrow{\nabla}\cdot\overrightarrow{\nabla}\to
  \overleftarrow{\partial}_kB^{kl}\overrightarrow{\partial}_l
\end{equation} 
must be made for the second-order derivative
operator in the dynamic action~\eqref{eq:J}.

In cases where the underlying microscopic system whose critical
behavior one is concerned with has cubic or orthorhombic lattice
symmetry, this matrix $\bm{B}$ is proportional to the unity matrix
$\bm{1}$ or at least diagonal, but for monoclinic and triclinic lattices it is
generally nondiagonal.  It can be transformed to
$\bm{1}$ by combining an orthogonal transformation $\bm{O}$
with $\bm{O}^{-1}\cdot\bm{B}\cdot\bm{O}=\text{diag}(b^1,\dotsc,b^d)\equiv \bm{b}$
to principal axes with a rescaling of coordinates. Let us make the
coordinate transformation
\begin{eqnarray}  \label{eq:transf}
\varphi: \bm{x}=(x^k)\mapsto\bm{x}'&=& \left(x^{\prime k}=\varphi^k(x^1,\dotsc,x^d)\right)\nonumber\\
&=&\bm{b}^{-1/2}\cdot\bm{O}^{-1}\cdot\bm{x}
\end{eqnarray}
and introduce the transformed quantities
\begin{eqnarray}
  \label{eq:primedpar}
\bm{\phi}'(\bm{x}',t)&=& B^{1/4}\,\bm{\phi}(\bm{x},t)\;,\nonumber\\
    \tilde{\bm{\phi}}'(\bm{x}',t)&=&B^{1/4}\, 
\tilde{\bm{\phi}}(\bm{x},t)\;,\nonumber\\ 
\mathring{\tau}'&=&\mathring{\tau}\;,\nonumber\\
\mathring{u}'&=&B^{-1/2}\,\mathring{u}\;,\nonumber\\
\mathring{h}'(\bm{x}',t)&=&B^{1/4}\,\mathring{h}(\bm{x},t)\;,\nonumber\\
\mathring{\lambda}'&=&\mathring{\lambda}\;,\nonumber\\
\mathfrak{V}'&=&\varphi(\mathfrak{V})\;,
\end{eqnarray}
where $B\equiv\det\bm{B}$.

Consider the analogs of the Hamiltonian~\eqref{eq:Ham} and dynamic
action~\eqref{eq:J} with the modified square gradient terms
Eqs.~\eqref{eq:modsgt} and \eqref{eq:repnablanabla},
respectively. Using the definitions~\eqref{eq:primedpar}, we can express
their contributions involving the volume integrals
$\int_{\mathfrak{V}}d^dx$ in terms of transformed (``primed'')
quantities. One easily verifies that the resulting expressions 
are identical with the original ones given in Eqs.~\eqref{eq:Ham} and
\eqref{eq:J}, up to the replacement of unprimed by primed
quantities. In particular, the primed square gradient term takes the
standard form
\begin{equation}\label{eq:standardprimedsgt}
 \frac{1}{2}\int_{\mathfrak{V}'=\varphi(\mathfrak{V})}d^dx'\,
 \sum_{k=1}^d\frac{\partial\bm{\phi}'}{\partial x^{\prime k}}\cdot
 \frac{\partial\bm{\phi}'}{\partial x^{\prime k}}\;. 
\end{equation}

All those transformations given in Eq.~\eqref{eq:primedpar} that refer to static bulk quantities and the mapping of the finite-size region $\mathfrak{V}$ are consistent with
those of Ref.~\onlinecite{Doh08} (whose matrix $\bm{A}$ corresponds to
our $\bm{B}$). The remaining ones are required for dynamics. To cope with free boundaries, we should also determine which boundary contributions to the transformed Hamiltonian and dynamic action result from the boundary integrals $\int_{\mathfrak{B}_j}d^{d-1}y\ldots$ of Eqs.~\eqref{eq:Ham} and \eqref{eq:J}, respectively. Before we do this, let us briefly discuss what it means for bulk critical behavior that the Hamiltonian describing the asymptotic critical behavior of weakly anisotropic systems can be mapped onto a primed one with coefficient matrix $\bm{B}'=\bm{1}$. 

There is no question that this mapping to primed variables involves nonuniversal
parameters. It is also clear that for nondiagonal $B^{kl}$,
part of the nonuniversality resides in the directions of the principal
axes, as emphasized by Dohm.\cite{Doh08} Does this mean that fundamental concepts of the modern theory of critical phenomena such as the notion of universality classes for static bulk critical behavior must be questioned, crucially modified or even given up? We see no reasons for such a conclusion. In our view, the very existence of the above mapping to simple minimal models such as the conventional $\phi^4$ theory is a clear signature of universality because it ensures that the critical properties of weakly anisotropic system can be expressed in terms of the universal properties of the latter. (Inasmuch as the explicit results of Ref.~\onlinecite{Doh08} are concerned, we are not aware of discrepancies; yet, our view of the situation may not fully be shared by its author.)

The nonuniversal geometric dependences contained in the metric should come as no surprise.  An essential element of the modern theory of critical phenomena is a mapping of microscopic models onto conceptually simple continuum models such as the $\phi^4$ theory whose critical fixed points describe the respective universality classes of static bulk critical behavior. Such  mappings  always involve nonuniversal parameters. Two familiar examples of such parameters are the location of the critical point and the slope of the coexistence curve; their nonuniversality shows up in the dependence of the two relevant scaling fields $g_\tau$ and $g_h$ on $T-T_{c,\infty}$ and the deviation of the magnetic field or chemical potential $\mu$ from their critical values $H_{c,\infty}=0$ or $\mu_{c,\infty}$,\cite{RGrev} where it should be recalled that in the case of fluids, both the thermal and magnetic scaling fields are nontrivial linear combinations of $T-T_{c,\infty}$ and $\mu-\mu_{c,\infty}$, up to nonlinear contributions.

The nonuniversal geometric dependences contained in the metric are of an analogous kind.  Absorbing them through the choice of properly defined transformed quantities is similar to the adsorption of other nonuniversal properties such as the location of the critical point and the slope of coexistence curve via appropriately chosen scaling fields. Of course, in any comparison of predictions of the theory with experimental results or Monte Carlo simulations for systems with weak
anisotropy, the nonuniversal geometry associated with the metric must be taken into account since it enters the way lattice observables depend on the order parameter. 

In most studies of critical behavior either standard square gradient terms with $\bm{B}=\bm{1}$ are chosen from the outset or else it is tacitly assumed that the above  transformation to primed variables has been made. This is done, in particular, when two-scale-factor is defined and discussed. In Ref.~\onlinecite{Doh08} it is
emphasized that two-scale-factor universality is broken unless
$\bm{B}=\bm{1}$.  Formally, this is correct since nonuniversal parameters that cannot be absorbed in
the nonuniversal amplitudes of the two relevant scaling fields
$g_\tau$ and $g_h$ are involved.  However, we believe it is natural and more reasonable to define two-scale-factor universality only \emph{after} the transformation to primed variables has been made. How the two-scale-factor universality of
the  corresponding  conventional $\phi^{\prime 4}$ theory manifests itself in
the original one with non-Euclidean metric follows from the relation between these two theories (as is worked out in detail for the case of static critical behavior in Ref.~\onlinecite{Doh08}). 

These considerations generalize in a straightforward fashion to the case of dynamic bulk critical behavior described by model $A$ with $\bm{B}\ne \bm{1}$. To analyze the corresponding asymptotic dynamic critical behavior, a dynamic scaling field (associated with the Onsager coefficient $\lambda$) must be considered in addition to $g_\tau$ and $g_h$. Hence a third nonuniversal scale factor must be fixed in the corresponding transformed theory with Euclidean metric.

A first obvious, though important, new feature one encounters when extending these considerations to finite-size systems is that the integration region $\mathfrak{V}$ and its boundary $\partial\mathfrak{V}$ transform under the map $\varphi$.  (Note that momenta would
transform with the inverse map $\varphi^{-1}$, so integration regions in
momentum space and hence momentum cutoffs would transform as well.) For general matrices $\bm{B}$ this is a shear transformation; cubes of finite linear dimension get transformed into parallelepipeds. 
Therefore, finite-size systems of a given (say, cubical) shape that involve the generalized square gradient terms~\eqref{eq:modsgt} and \eqref{eq:repnablanabla} should not be compared to their primed analogs of the same but of a different (non-cubical) shape. 

According to the phenomenological theory of finite-size scaling, the finite-size critical behavior of a given microscopic system does not only depend on those properties that determine the corresponding bulk universality class but also on gross finite-size properties such as shape and boundary conditions. Hence, each universality class for static bulk critical behavior generally splits up into several universality classes for finite-size critical behavior. This is analogous to the splitting up of static bulk universality classes into several distinct universality classes for dynamic bulk critical behavior\cite{HH77} and into those for static boundary critical behavior.\cite{Die86a,Die97} The upshot is that two  finite-size systems with the same volume region $\mathfrak{V}$ (and hence shape) whose large-length scale descriptions involve generalized and standard square gradient terms, respectively, may represent distinct finite-size universality classes even when the same kinds of  boundary conditions (e.g., periodic boundary conditions) are chosen for both of them on the level of lattice models. 

Square gradient terms with nondiagonal $\bm{B}$ give rise to important modifications of the finite-size critical behavior of systems that are finite in all directions. This is discussed in detail for the case of static critical phenomena in Ref.~\onlinecite{Doh08}. In the case of our slab geometry, the image $\mathfrak{V}'$ of the slab of infinite lateral extension and thickness $L$ under $\varphi$  is again a slab whose thickness $L'$ generally differs from $L$. However, we must also clarify how the boundary conditions are affected by generalized gradient terms and what happens to them under the mapping to the primed system. Considering a Hamiltonian and dynamic action that agree with those specified by Eqs.~\eqref{eq:Ham} and \eqref{eq:J} except for the replacements of $(\nabla\bm{\phi})^2/2$ and $\overleftarrow{\nabla}\cdot\overrightarrow{\nabla}$  by the generalized gradient term~\eqref{eq:modsgt} and the operator~\eqref{eq:repnablanabla}, respectively, one finds that the boundary conditions now become
\begin{eqnarray}\label{eq:modbc}
       \bm{n}^{\text{T}}\cdot\bm{B}\cdot\nabla\tilde{\bm{\phi}}(\bm{x},t)&=&
  \mathring{c}_j\,\tilde{\bm{\phi}}(\bm{x},t)\,,
  \;\;\bm{x}\in\mathfrak{B}_j\,,\nonumber \\[\smallskipamount]
  \bm{n}^{\text{T}}\cdot\bm{B}\cdot\nabla\bm{\phi}(\bm{x},t)&=&\mathring{c}_j\,\bm{\phi}(\bm{x},t)\,,
\;\;\bm{x}\in\mathfrak{B}_j\,,
\end{eqnarray}
where $\bm{n}^{\text{T}}$ is the row vector transposed to the column vector $\bm{n}$. 

Upon introducing the vector $\bm{f}=f\hat{\bm{f}}=\bm{B}\cdot\bm{n}$, one recognizes the derivatives on the left-hand sides of these equations as directional derivatives $\partial_{\bm{f}}\bm{\phi}=f\partial_{\hat{\bm{f}}}\bm{\phi}$. Recall that in our case $\bm{n}=\pm\bm{e}_z$  on $\mathfrak{B}_{1,2}$. For general $\bm{B}$, the vector $\bm{f}$ is not parallel to $\bm{n}$. The  meaning of these boundary conditions can be understood as follows. Suppose we extrapolate the fields  $\bm{\phi}$ and $\tilde{\bm{\phi}}$  from a point $\bm{x}_{\mathfrak{B}_j}$ on boundary plane $\mathfrak{B}_j$ along the direction $-\hat{\bm{f}}$ in a linear fashion. Then these extrapolations vanish  when the coordinate differences $(\bm{x}-\bm{x}_{\mathfrak{B}_j})\cdot \hat{\bm{f}}$ take the values $\pm f/\mathring{c}_j$ for $j=1$ and $2$, respectively.  Thus, the modified square gradient term in general  changes both the direction of the extrapolation and the associated extrapolation length. Further, the vector $\bm{f}$ is normal to the surface $\mathfrak{B}$ in the metric $(B_{kl})$. To see this, let us represent vectors $\bm{v}$ and $\bm{w}$ in the canonical basis $\{\bm{e}_j\}$ of $\mathbb{R}^d$ as $\bm{f}=f^j\bm{e}_j$ and denote the scalar product in the metric $(B_{kl})$ as $B(\bm{v},\bm{w})=B_{kl}v^kw^l$. The dual basis $\{\bm{f}^{(k)}\}$, which satisfies $B(\bm{f}^{(k)},\bm{e}_l)={\delta^k}_l$, is given by $\bm{f}^{(k)}=B^{kl}\bm{e}_l$. For points on $\mathfrak{B}_{1,2}$, the vector $\bm{f}$ is nothing but $\pm\bm{f}^{(d)}$, and hence orthogonal to the tangent vectors $\bm{e}_{k\ne d}$ in the metric $(B_{kl})$. In fact, the directional derivative $\partial_{\bm{f}}$ on the left-hand sides of Eq.~\eqref{eq:modbc} corresponds to a  normal derivative in this metric. Transforming to primed variables,  
gives the Robin boundary conditions $\partial'_n\bm{\phi}'=\mathring{c}_j'\bm{\phi}'$ for the fields $\bm{\phi}'$ and $\tilde{\bm{\phi}}'$ with $\mathring{c}_j'=\mathring{c}_j/f$, in conformity with our above results.

Since the enhancements variables $\mathring{c}_j$ and $\mathring{c}_j'$ of the systems with generalized and standard square gradients are proportional to each other, the fixed points of the transformed system with $c'_j=0,\pm\infty$  map onto the respective  fixed points of the unprimed system; the effects of the generalized square gradient terms implied by weak anisotropy normally may be expected  to be of a purely geometrical kind. They should be particularly important and interesting when $\bm{B}$ is nondiagonal.

One class of systems deserving detailed studies consists of binary alloys. This is because their description generally involves, besides the order parameter, further, so-called secondary densities (nonordering densities). Studies of the static boundary critical behavior of body centered cubic binary alloys have revealed that careful investigations of the coupling of the order parameter to these secondary densities and the symmetry reduction caused by the presence of boundary planes may be necessary to determine which surface universality class applies. In fact, depending on the orientation of the surface plane relative to the crystal axes, distinct surface universality classes may be realized.\cite{DLBD97LD98,KDN+97} Analogous studies have yet to be performed for binary alloys with less symmetric (e.g., monoclinic and triclinic) crystal structures, which would yield nondiagonal metrical coefficients $B^{kl}$.

We close with a brief discussion of an elementary example of a slab exhibiting weakly anisotropic critical behavior.  Consider a nearest-neighbor (NN) lattice $O(n)$ spin model that is restricted to the layers $z=0,1,\dotsc,L$ of the simple cubic lattice $\mathbb{Z}^d$. To introduce weak anisotropy, we assume that all NN  bonds perpendicular to the bottom and top layers $z=0$ and $z=L$ have strength $J_{\perp}$, while those along the top, bottom, and remaining layers have different strengths $J_{\|,1}$, $J_{\|,2}$, $J_{\|}$, respectively. Mapping this microscopic model onto a continuum model gives squared gradient terms with $\bm{B}=\text{diag}(b^y,\dotsc,b^y,b^z)$, where $b^z/b^y=J_\perp/J_\|$. Hence the ratio of the corresponding bulk correlation lengths $\xi_{\infty,\|}$ and $\xi_{\infty,\perp}$ (defined via second moments of the respective displacements parallel and perpendicular to the layers $z=\text{const}$), satisfies $\xi_{\infty,\|}/\xi_{\infty,\perp}=(b^y/b^z)^{-1/2}$. The rescaling $z'=(b^z/b^y)^{-1/2}z$ maps the large-scale continuum field theory of this film on a primed system with $\bm{B}\propto \bm{1}$ and film thickness $L'=(b^z/b^y)^{-1/2}L$. Writing the $L$-dependent part of the  finite-size free energy per cross-sectional area $L^{d-1}$ and $k_BT$ at the bulk critical point as $\Delta_{\text{ai}}^{(\wp)}/L^{d-1}$, we can introduce a Casimir amplitude $\Delta_{\text{ai}}^{(\wp)}$ for the weakly anisotropic system, where $\wp$ indicates which  (scale-invariant) boundary conditions hold on sufficiently large scales. This could be anyone of those considered in Ref.~\onlinecite{KD92a}, namely  periodic, antiperiodic, Dirichlet-Dirichlet, Dirichlet-special, and special-special), as well as $++$, $--$, and $+-$ boundary conditions. Upon expressing $\Delta_{\text{ai}}^{(\wp)}/L^{d-1}$ in terms of $L'$, we see that $\Delta_{\text{ai}}^{(\wp)}$ is related to the Casimir amplitude $\Delta^{(\wp)}$ of the transformed (isotropic) system via 
\begin{equation}\label{eq:Dai}
\Delta_{\text{ai}}^{(\wp)}=(J_\|/J_\perp)^{-(d-1)/2}\,\Delta^{(\wp)}\;.
\end{equation}

This relation was obtained for the special case of antiperiodic boundary conditions (AP) in a recent paper by Dantchev and Gr\"uneberg\cite{DG08b}  who determined $\Delta_{\text{ai}}^{(\text{AP})}$  in  the large-$n$ limit by the exact solution of a mean spherical model. Our reasoning shows that it holds more generally and  follows from simple considerations.

\begin{acknowledgments} 
  This work was completed while one of us (H.W.D.) participated in the
  KITP research program ``The theory and practice of
  fluctuation-induced interactions.'' He would like to thank the
  organizers for their kind invitation and the stimulating atmosphere,
  and the Kavli Institute for Theoretical Physics (KITP) for their
  hospitality and support. He is grateful to Daniel Gr\"uneberg for
  discussions and critical reading of the manuscript, to Ron Horgan
  for calling his attention to Refs.~\onlinecite{DHLR98} and a
  discussion, and to Felix Schmidt for checking part of the
  calculations in Appendix C. We also thank an anonymous referee for
  requesting us to comment on possible universality violations and
  weak anisotropy effects of the sort discussed in
  Refs.~\onlinecite{CD02b} and \onlinecite{Doh08}.

  This research was supported in part by the Deutsche
  Forschungsgemeinschaft under Grant No.\ Di-378/5 (H.W.D.), the National
  Science Foundation under Grant No.\ PHY05-51164 (H.W.D.), and by the
  Bulgarian Fund for Scientific Research under Grant No F-1517 (H.C.).
  We gratefully acknowledge the support of all three funding agencies.
\end{acknowledgments}
\appendix

\section{Representations and conventions}\label{sec:conv}
We define Fourier transforms with respect to time and the position
vector's component $\bm{y}$ along the film by
\begin{eqnarray}
  \label{eq:FTpom}
  \lefteqn{\bm{\phi}(\bm{x},t)\equiv\bm{\phi}(\bm{y},z,t)=\int_\omega
  \bm{\phi}(\bm{y},z,\omega)\,e^{-i\omega t}}&& \nonumber\\
  &=&\int_{\bm{p}}\bm{\phi}(\bm{p},z,t)\,e^{i\bm{p}\cdot\bm{y}}=
  \int_{\omega,\bm{p}}\bm{\phi}(\bm{p},z,\omega)\,
  e^{i(\bm{p}\cdot\bm{y}-\omega t)}\,,\quad 
\end{eqnarray}
where we employ the short-hand notations
\begin{equation}
  \label{eq:intcon}
  \int_\omega=\int_{-\infty}^\infty \frac{d\omega}{2\pi}\;, \quad 
  \int_{\bm{p}}=\int_{\mathbb{R}^{d-1}}
  \frac{d^{d-1}p}{(2\pi)^{d-1}}\,.  
\end{equation}
Note that we do not introduce separate symbols for a function such as
$\bm{\phi}(\bm{x},t)$ and its Fourier transforms
$\bm{\phi}(\bm{p},z,t)$, $\bm{\phi}(\bm{p},z,\omega)$, and
$\bm{\phi}(\bm{p},z,\omega)$; which quantity is meant should be clear
from the arguments of these quantities.

When defining Fourier transforms of multi-point response and cumulant
functions that are invariant under translations $\bm{y}_i\to
\bm{y}_i+\bm{y}_0$ parallel to the boundary planes and time
translations $t_i\to t_i+t_0$, we mean by the respective Fourier
transforms the coefficients of the momentum and frequency conserving
factors $(2\pi)^{d-1}\,\delta(\sum_i\bm{p}_i)$ and
$2\pi\,\delta(\sum_i\omega_i)$, respectively. For example, the Fourier
transforms $R(\bm{x};\bm{x}';\omega)$ and $R(\bm{p};z,z';\omega)$ of
the free response propagator $R(\bm{x},t;\bm{x}',t')$ satisfy the
relations 
\begin{eqnarray}
  \label{eq:RFT}
  \lefteqn{R(\bm{x},t;\bm{x}',t')=
    \int_{\omega}R(\bm{x};\bm{x}';\omega) \,e^{-i\omega(t-t')}}&&
  \nonumber\\ 
&=&\int_{\omega,\bm{p}}R(\bm{p};z,z';\omega)\,
e^{-i\omega(t-t')}\,e^{i\bm{p}\cdot(\bm{y}-\bm{y}')} \,.
\end{eqnarray}

\section{Free response propagator}
\label{sec:frp}

To determine the free response propagator $R_L(\bm{p};z,z';\omega)$ for
general non-negative values of $\mathring{c}_1$, $\mathring{c}_2$, and $\mathring{\tau}$, we must
solve the differential equation
\begin{equation}
  \label{eq:SLdgl}
  [-i\omega+\mathring{\lambda}(\mathring{\tau}+p^2-\partial_z^2)]R_L(\bm{p};z,z';\omega)=\delta(z-z')
\end{equation}
with the boundary conditions~\eqref{eq:fmbc}.  Two linearly
independent solutions of this Sturm-Liouville differential equation are
$e^{\pm\mathring{\kappa}_\omega z}$, where $
  \mathring{\kappa}_\omega$ was defined in Eq.~\eqref{eq:kapom}.
From them we can construct the two linear combinations
\begin{equation}
  \label{eq:U1}
  U_1(z)= \mathring{\kappa}_\omega \cosh(\mathring{\kappa}_\omega z)+\mathring{c}_1\sinh(\mathring{\kappa}_\omega z)
\end{equation}
and
\begin{equation}
  \label{eq:U2}
  U_2(z)= \mathring{\kappa}_\omega
  \cosh[\mathring{\kappa}_\omega(L-z)]+\mathring{c}_2\sinh[\mathring{\kappa}_\omega(L- z)]
\end{equation}
that fulfill the boundary conditions on the boundary planes $z=0$ and
$z=L$, respectively. The Green's function
$\mathring{\lambda}\,R_L(\bm{p};z,z';\omega)$ is given by
$-U_1(z_<)\,U_2(z_>)/W_{12}$, where $z_<=\min(z,z')$ and
$z_>=\max(z,z')$.\cite{CH68} The normalization constant $W_{12}$ is
fixed by the jump condition $\mathring{\lambda} [\partial_z
R_L]^{z=z'+0}_{z=z'-0}=-1$. This yields the Wronskian
  \begin{equation}
    \label{eq:Wronskian}
    W_{12}=
    \begin{vmatrix}
      U_1(z')&U_2(z')\\
U_1'(z')&U_2'(z')
    \end{vmatrix}\;,
\end{equation}
which in our case is independent of $z'$. A straightforward calculation
then gives
\begin{widetext}
\begin{eqnarray}
  \label{eq:Gfree}
  \lefteqn{\mathring{\lambda}\,R_L(\bm{p};z,z';\omega|\mathring{c}_1,\mathring{c}_2)}&&\nonumber \\ 
&=&\theta(z'-z)\,
\frac{[\mathring{c}_1 \sinh(\mathring{\kappa}_\omega z)+\mathring{\kappa}_\omega\cosh
   (\mathring{\kappa}_\omega z)  ]
   [\mathring{c}_2 \sinh [\mathring{\kappa}_\omega(L-z'
   )]+\mathring{\kappa}_\omega\cosh[\mathring{\kappa}_\omega(L-z' )]}{\mathring{\kappa}_\omega^2(\mathring{c}_1+\mathring{c}_2) \cosh
   (\mathring{\kappa}_\omega L) +\mathring{\kappa}_\omega (\mathring{\kappa}_\omega^2+\mathring{c}_1
   \mathring{c}_2)\sinh(\mathring{\kappa}_\omega L ) }
\nonumber\\ &&\strut
+(z\leftrightarrow z'),
\end{eqnarray}
where we have explicitly indicated the enhancement variables $\mathring{c}_j$
on the left-hand side for clarity. In the special case
$\mathring{c}_1=\mathring{c}_2=0$, this simplifies to
\begin{equation}
  \label{eq:G00}
\mathring{\lambda}\,R_L(\bm{p};z,z';\omega|0,0)=\theta(z'-z)\,
  \frac{\cosh(\mathring{\kappa}_\omega z) \cosh[\mathring{\kappa}_\omega(L-z'
    )]}{\mathring{\kappa}_\omega  \sinh(\mathring{\kappa}_\omega L)}+(z\leftrightarrow z')\;.
\end{equation}

The free response propagator for periodic boundary conditions can be
determined in a straightforward fashion by performing the summation in
Eq.~\eqref{eq:RLperir} using Eq.~\eqref{eq:kapom}. This gives
\begin{equation}
  \label{eq:Rper}
\mathring{\lambda}\,R^{(\text{per})}_L(\bm{p};z-z';\omega)=
\frac{1}{2\kappa_\omega}\,
  \frac{\cosh[\kappa_\omega(L/2-|z-z'|)]}{\sinh(\kappa_\omega L/2)}\,.
\end{equation}
\end{widetext}

\section{Calculation of Feynman graphs}
\label{sec:graphs}

\subsection{Layer susceptibility for periodic boundary conditions}
\label{sec:cofpbc}

The first one-loop graph of $\chi_{zz}$ shown in
Eq.~\eqref{eq:chizzpt} involves the static propagator
$G_{L,\psi}^{(\text{per})}(\bm{x}|\mathring{\tau})=
\mathring{\lambda}\,R^{(\text{per})}_{L,\psi}(\bm{x},\omega=0|\mathring{\tau})$ at zero
separation $\bm{x}=\bm{0}$.  To compute it at $\mathring{\tau}=0$, we add and
subtract the zero-mode contribution. Since this vanishes (in
dimensional regularization) when $\mathring{\tau}=0$, we may simply replace
$G_{L,\psi}^{(\text{per})}(\bm{x}|0)$ by
$G_{L}^{(\text{per})}(\bm{x}|0)$.  It is then convenient to use the
latter's representation~\eqref{eq:RLperir} in terms of image
contributions, with $G^{(d)}_\infty(\bm{x}|0)= 2^{-2}\pi^{-d/2}\,
\Gamma(d/2-1) \, x^{2-d}$. The term with $m=0$ vanishes since
$G^{(d)}_\infty(\bm{0}|0)=0$ in dimensional regularization. The
summation of the remaining terms is straightforward, giving
\begin{eqnarray}
  \label{eq:Epertauzero}
  G^{(\text{per})}_L(\bm{x}=\bm{0}|\mathring{\tau}=0)=
  \frac{\Gamma(d/2-1)}{2\,\pi^{d/2}\, L^{d-2}}\,\zeta(d-2)\;.
\end{eqnarray}

Since the ${k=0}$ component of this quantity vanishes when $\mathring{\tau}=0$, we
can directly use this result to compute the one-loop graph with the
dashed blue line of Eq.~\eqref{eq:chizzpt}. Upon performing the required
$z$-integration of the external legs
$R_L^{(\text{per})}(\bm{p},z,\omega|0)^2$, one obtains 
\begin{eqnarray}
  \label{eq:oneloopchizzpsi}
  \mathring{\lambda}\;\raisebox{-3mm}{\includegraphics[scale=0.8]{oneloopchizzpsi.eps}}&=&
 -\frac{n+2}{3}\,\frac{\mathring{u}}{2}\,
 \frac{\Gamma(d/2-1)\,\zeta(d-2)}{2\pi^{d/2}\,L^{d-5}}\nonumber\\ &&\times
 \frac{\mathring{\kappa}_\omega L +\sinh(L\mathring{\kappa}_\omega)}{
   8\mathring{\kappa}_\omega^3L^3\sinh^2(\mathring{\kappa}_\omega L/2)} \;.
\end{eqnarray}

The remaining graph of  Eq.~\eqref{eq:chizzpt} involves the dotted red
line, which is
$L^{-1}\,G_\infty^{(d-1)}(\bm{y}=\bm{0}|\delta\mathring{\tau}_L^{(\text{per})})$ in
position space. Hence we have
\begin{eqnarray}
  \label{eq:oneloopchizzvarphi}
  \mathring{\lambda}\;
  \raisebox{-3mm}{\includegraphics[scale=0.8]{oneloopchizzvarphi.eps}}&=& 
\frac{n+2}{3}\,\frac{\mathring{u}}{2}\,
A_{d-1}\,L^2[\delta\mathring{\tau}^{(\text{per})}_L]^{(d-3)/2}\nonumber\\ &&\times
\frac{\mathring{\kappa}_\omega L +\sinh(\mathring{\kappa}_\omega L)}{
  8\mathring{\kappa}_\omega^3 L^3\sinh^2(\mathring{\kappa}_\omega L/2)} \;,
\end{eqnarray}
where $\delta\mathring{\tau}^{(\text{per})}_L$ is the $O(\mathring{u})$ shift given in
Eq.~\eqref{eq:shifts}. 

\subsection{Surface susceptibilities for sp-sp boundary conditions}
\label{sec:spspbc}

The analogs of the graph~\eqref{eq:oneloopchizzpsi} for
$\chi^{(\text{sp-sp})}_{11}$ and $\chi^{(\text{sp-sp})}_{1L}$ involve
the static propagator $G_{L,\psi}^{(\text{N-N})}(\bm{x},\bm{x}'|0)
=\mathring{\lambda}\,R^{(\text{N-N})}_{L,\psi}(\bm{x},\bm{x}',{\omega=0}|0)$ at equal
positions $\bm{x}=\bm{x}'$. Since its zero-mode contribution vanishes
at $\mathring{\tau}=0$, it agrees with $G_{L}^{(\text{N-N})}(\bm{x},\bm{x}|0)$. We
use its
representation~\eqref{eq:RLDDNNir} and take into account that  the
$m=0$ contribution of the first sum vanishes at $\mathring{\tau}=0$. The
remaining terms can be summed in a straightforward fashion to obtain
\begin{eqnarray}
  \label{eq:Esptau0}
  G_L^{(\text{N-N})}(\bm{x},\bm{x}|0)&=&
  \frac{\Gamma(d/2-1)}{2^d\,\pi^{d/2}\,L^{d-2}}\, 
  [2\,\zeta(d-2)\nonumber \\
  &&\strut +\zeta(d-2,z/L)+\zeta(d-2,1-z/L)]\,.\nonumber \\
\end{eqnarray}

Note that the two generalized Hurwitz zeta functions\cite{rem:Hurwitz} $\zeta(2-\epsilon,s)$ and
$\zeta(2-\epsilon,1-s)$ behave as $s^{\epsilon-2}$ and
$(1-s)^{\epsilon-2}$ at small values of their arguments $s=z/L$ and
$1-s$, respectively. They contain the uv singular contributions
$-\epsilon^{-1}\delta'(s)$ and $-\epsilon^{-1}\delta'(1-s)$.

To show this more clearly and to determine the Laurent expansion of
the graph in question to order $\epsilon^0$, we proceed as follows. We
transform to the variable $s$. The $z$-independent part of
Eq.~\eqref{eq:Esptau0} leads to contributions of the graphs that can
be expressed in term of the integrals
\begin{equation}
  \label{eq:Ijk}
  I_{j,k}(\mathring{\kappa}_\omega L)=\int_0^1ds\,f_j(s;\mathring{\kappa}_\omega L)\,f_k(s;\mathring{\kappa}_\omega L)\;,\;\;j,k=1,2\,,
\end{equation}
with
\begin{equation}
  \label{eq:fj}
  f_1(s;\varkappa)=\frac{\cosh [(1-s) \varkappa]}{\varkappa\sinh \varkappa  }
\end{equation}
and
\begin{equation}
  \label{eq:f2}
  f_2(s;\varkappa)=f_1(1-s;\varkappa)\;,
\end{equation}
where  $Lf_j(z/L;\mathring{\kappa}_\omega L)/\mathring{\lambda} =R_L^{(\text{N-N})}(\bm{p};z,(j-1)L;\omega|0)$
represents an external leg attached to $\mathfrak{B}_j$. 

The required integrations are elementary, giving
\begin{equation}
  \label{eq:Ijjres}
  I_{1,1}(\varkappa)=I_{2,2}(\varkappa)=\frac{\sinh(2\varkappa)+2\varkappa}{4\varkappa^3
    \sinh^2\varkappa}
\end{equation}
and
\begin{equation}
  \label{eq:I12res}
  I_{1,2}(\varkappa)=I_{2,1}(\varkappa)=\frac{1+\varkappa\coth \varkappa}{2\varkappa^3\sinh \varkappa}\;.
\end{equation}

The contribution produced by the term $\propto\zeta(2-\epsilon,s)$ in
Eq.~\eqref{eq:Esptau0} involves the integrals  
\begin{equation}
  \label{eq:Jjk}
  J_{j,k}(\epsilon;\varkappa)=\int_0^1ds\,\zeta(2-\epsilon,s)\,\,f_j(s;\varkappa)\,f_k(s;\varkappa)
\end{equation}
with $(j,k)=(1,1)$ and $(1,2)$. The analogous contribution implied by
the term  $\propto\zeta(2-\epsilon,1-s)$ can also be expressed in
terms of these integrals, as can be seen  by making a change of
variables $1-s\to s$ and using Eq.~\eqref{eq:f2}.

To compute the Laurent expansion of these integrals, we substitute
\begin{equation}
  \label{eq:zetaid}
  \zeta(2-\epsilon,s)=s^{\epsilon-2}+\zeta(2-\epsilon,s+1)
\end{equation}
and decompose each one of them into a sum of the respective two integrals
\begin{equation}
  \label{eq:Sjk}
  S_{j,k}(\epsilon;\varkappa)=\int_0^\infty ds\,s^{\epsilon-2}\,f_j(s;\varkappa)\,f_k(s;\varkappa)\,\theta(1-s)
\end{equation}
and 
\begin{equation}
  \label{eq:Rjk}
  R_{j,k}(\epsilon;\varkappa)=\int_0^1ds\,\zeta(2-\epsilon,s+1)\,f_j(s;\varkappa)\,f_k(s;\varkappa)\;.
\end{equation}

The latter integrals are regular at $\epsilon=0$, and hence can be
expanded as
$R_{j,k}(\epsilon;\varkappa)=R_{j,k}(0;\varkappa)+O(\epsilon)$.  The
integrals $S_{j,k}(\epsilon;\varkappa)$ may be viewed as the results
of applying the distribution denoted $s_+^{\epsilon-2}$ in
Ref.~\onlinecite{GS64} to the test functions
$h(s)=f_j(s;\varkappa)\,f_k(s;\varkappa)\,\theta(1-s)$. The Laurent
expansion of this distribution is well known.\cite{GS64} It reads
\begin{equation}
  \label{eq:Laurentdist}
  s_+^{\epsilon-2}=\frac{-1}{\epsilon}\,\delta'(s)+s_+^{-2}+O(\epsilon)\,,
\end{equation}
where the generalized function $s_+^{-2}$ is defined by (cf.\ the
Appendix of Ref.~\onlinecite{Die86a})
\begin{eqnarray}
  \label{eq:splus}
  \int_{-\infty}^\infty
  ds\,s_+^{-2}\,h(s)&=&\int_0^1ds\,s^{-2}\,[h(s)-h(0)-s\,h'(0)]
  \nonumber\\ &&\strut +\int_1^\infty
  ds\,s^{-2}\,[h(s)-h(0)]\,.  
\end{eqnarray}

Utilizing these results, one arrives at
\begin{equation}
  \label{eq:Sjkexp}
  S_{j,k}(\epsilon;\varkappa)=S^{(-1)}_{j,k}(\varkappa)\,\epsilon^{-1}+S^{(0)}_{j,k}(\varkappa)+O(\epsilon)
\end{equation}
with 
\begin{eqnarray}
  \label{eq:PTSjk}
  S^{(-1)}_{j,k}(\varkappa)&=&f'_j(0;\varkappa)\,f_k(0;\varkappa)+f_j(0;\varkappa)\,f_k'(0;\varkappa)\nonumber\\
&=&-\frac{1}{\varkappa}
\begin{cases}
  2\coth(\varkappa)\,,&j=k=1\;,\\
\text{csch}(\varkappa)\,,&j\ne k\;,\\
0\,,&j=k=2\;,
\end{cases}
\end{eqnarray}
and
\begin{eqnarray}
  \label{eq:Sjk0}
 S^{(0)}_{j,k}(\varkappa)&=&
\int_0^1
  \frac{ds}{s^2}\,\big\{f_j(s;\varkappa)\,f_k(s;\varkappa)-f_j(0;\varkappa)\,f_k(0;\varkappa)\nonumber\\ &&\strut -s\,[f'_j(0;\varkappa)\,f_k(0;\varkappa)+f_j(0;\varkappa)\,f_k'(0;\varkappa)]\big\}\nonumber \\ &&\strut
 -f_j(0;\varkappa)\,f_k(0;\varkappa)\,,
\end{eqnarray}
where the prime on $f_j'(s;\varkappa)$ means a  derivative with respect
to $s$.

The latter integrals can be performed in a straightforward fashion
using {\sc Mathematica}.\cite{Mat} This yields
\begin{eqnarray}
    \label{eq:S110}
    S_{11}^{(0)}(\varkappa)&=&[\gamma_E -\text{Chi}(2 \varkappa )+\ln(2\varkappa )
      -1]\,\frac{2\coth \varkappa}{\varkappa}
    \nonumber\\ &&\strut + \frac{\varkappa\,\cosh (2 \varkappa )\, \text{Shi}(2 \varkappa
   )-1}{\varkappa^2\sinh^2\varkappa}\;,
  \end{eqnarray}
  \begin{eqnarray}
    \label{eq:S120}
S_{12}^{(0)}(\varkappa)&=&
    \frac{\gamma_E-\text{Chi}(2\varkappa  )+\ln (2\varkappa ) -1}{\varkappa\sinh \varkappa}\nonumber\\
  &&\strut  +\frac{\varkappa\,\text{Shi}(2\varkappa )-1}{\varkappa^2\sinh(\varkappa)\tanh\varkappa}\;,
  \end{eqnarray}
and
 \begin{equation}
    \label{eq:S22def}
S_{22}^{(0)}(\varkappa)=
   \frac{\varkappa\,\text{Shi}(2\varkappa)
     -\cosh^2\varkappa}{\varkappa^2 \sinh^2\varkappa}\;, 
  \end{equation}
where $\text{Chi}$ and $\text{Shi}$ are the hyperbolic cosine and sine
integrals, respectively. 

It is useful to introduce the combinations
\begin{eqnarray}
\label{eq:Sdef}
S(\varkappa)&\equiv&S_{1,1}^{(0)}(\varkappa)+
    S_{2,2}^{(0)}(\varkappa)\nonumber\\ &=&
    \frac{2 \coth (\kappa )}{\kappa }\big[ \gamma_E -1-\text{Chi}(2
    \kappa )+\ln (2 \kappa )\nonumber\\ && \strut+\coth (\kappa )
   \text{Shi}(2 \kappa )]-\frac{1+2\,\text{csch}^2\varkappa}{\varkappa^2}\,,
  \end{eqnarray}
  \begin{eqnarray}
    \label{eq:Rdef}
R(\varkappa)&=&R_{1,1}(0;\varkappa)+R_{2,2}(0;\varkappa)\nonumber\\ &=&
    \frac{\text{csch}^2\varkappa}{2\varkappa^2}\int_0^1ds\,\psi^{(1)}(s+1)\{2+\cosh(2s\varkappa)\nonumber\\ &&
\strut +\cosh [2(1-s) \varkappa ]\}
\,,
\end{eqnarray}
and the function
\begin{eqnarray}
  \label{eq:R2}
  R_2(\varkappa)&\equiv& 2R_{1,2}(0;\varkappa)\nonumber\\ &=&
\frac{2\,\text{csch}^2\varkappa}{\varkappa^2}\int_0^1ds\,\big\{
\psi^{(1)}(s+1)\nonumber\\ && \times  \cosh (s
\varkappa ) \cosh[ (1 -s) \varkappa]\big\}\,.\phantom{(C24)}
\end{eqnarray}
Then our results for the graphs involving the  dashed blue line can be
written as
\begin{eqnarray}
  \label{eq:oneloopchi11psi}
  \lefteqn{\mathring{\lambda}\;\raisebox{-3mm}{\includegraphics[scale=0.8]{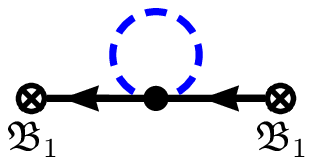}}}&& \nonumber\\ &=&
 -\frac{n+2}{3}\, \frac{\mathring{u}\mathring{\lambda}}{2}\,
 \frac{\Gamma(1-\epsilon/2)\,L^{1+\epsilon}}{(4\pi)^{2-\epsilon/2}}
 \bigg\{ -\frac{2\coth(\mathring{\kappa}_\omega
   L)}{\mathring{\kappa}_\omega L\,\epsilon} \nonumber\\ &&\strut
 +\frac{\pi^2}{3}\,I_{1,1}(\mathring{\kappa}_\omega L)  +
 S(\mathring{\kappa}_\omega L)  +
 R(\mathring{\kappa}_\omega L)+O(\epsilon)\bigg\} \phantom{(C25)}
\end{eqnarray}
and
\begin{eqnarray}
  \label{eq:oneloopchi12psi}
  \lefteqn{\mathring{\lambda}\;\raisebox{-3mm}{\includegraphics[scale=0.8]{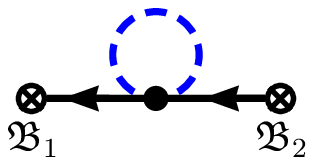}}}&&\nonumber\\ &=&
 -\frac{n+2}{3}\,\frac{\mathring{u}\mathring{\lambda}}{2}\,
 \frac{\Gamma(1-\epsilon/2)\,L^{1+\epsilon}}{(4\pi)^{2-\epsilon/2}}
 \bigg\{ -\frac{2\,\text{csch}(\mathring{\kappa}_\omega L)}{\mathring{\kappa}_\omega L\,\epsilon} 
\nonumber\\ && \strut
 +\frac{\pi^2}{3}\,I_{1,2}(\mathring{\kappa}_\omega L)
+2S_{1,2}^{(0)}(\mathring{\kappa}_\omega L)
+ R_2(\mathring{\kappa}_\omega L)+O(\epsilon)\bigg\}\,.\nonumber\\
\end{eqnarray}

The remaining graphs, involving the zero-mode propagator at the
shifted bare temperature $\delta\mathring{\tau}^{(\text{sp-sp})}_L$, are given by
\begin{eqnarray}
  \label{eq:oneloopchi11varphi}
\lefteqn{ \mathring{\lambda}\;
  \raisebox{-4mm}{\includegraphics[scale=0.8]{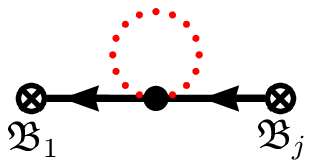}}}&&\nonumber\\ &=&
\frac{n+2}{3}\,\frac{\mathring{u}}{2}\,\frac{A_{d-1}}{L}\,
[\delta\mathring{\tau}_L^{(\text{sp-sp})}]^{(d-3)/2} \,
 I_{1,j}(\mathring{\kappa}_\omega L)\;.\phantom{(C27)}
\end{eqnarray}

While the latter graph is uv finite, the previous two contain pole
terms. Recalling the $O(u)$ result\cite{DD81b,DD83a,Die86a}
$Z_1=1+u\,(n+2)/{3\epsilon}+O(u^2)$, one sees that they cancel with the
contribution $(Z^{-1}_1Z^{-1}_\phi-1)\lambda
\mu^{-4}\,\raisebox{-0.9em}{\includegraphics[width=4em]{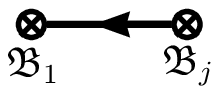}}$.

In our calculation of the zero-momentum limits of the inverse
{$\omega=0$}~susceptibilities $\chi_{11,R}^{-1}$ and $\chi_{1L,R}^{-1}$,
the asymptotic behaviors of the functions $S$, $R$, $R_2$, and
$S^{(0)}_{12}$ as $\varkappa\to 0$ are needed. Using the known 
limiting forms
$\text{Chi}(\varkappa)=\gamma_E+\ln\varkappa+O(\varkappa^2)$ and
$\text{Shi}(\varkappa)=\varkappa+O(\varkappa^3)$, one finds that
\begin{equation}
  \label{eq:Sassmall}
  S(\varkappa)=2S^{(0)}_{12}(\varkappa)+O(\varkappa^{-2})=-2\varkappa^{-4}+O(\varkappa^{-2})
\end{equation}
for small $\varkappa$. To determine the asymptotic forms of the
functions $R(\varkappa)$ and $R_2(\varkappa)$, we expand the integrands
in the integrals of Eqs.~\eqref{eq:Rdef} and \eqref{eq:R2} in powers
of $\varkappa$. Performing the $s$-integral for the lowest-order term
and expanding the prefactors leads to
\begin{equation}
  \label{eq:Rassmall}
      R(\varkappa)=
R_2(\varkappa)+O(\varkappa^{-2})=
    2\varkappa^{-4}+O(\varkappa^{-2})\;.
\end{equation}


\begin{thebibliography}{80}
\bibitem[{rem({\natexlab{a}})}]{remFSE}
\bibinfo{note}{The application of the field-theoretic RG to the study of
  finite-size effects started with Symanzik's seminal paper \cite{Sym81} on
  systems confined between parallel plates under Dirichlet boundary conditions,
  Br{\'e}zin's work on the large-$n$ limit, \cite{Bre82} and two papers
  \cite{BZJ85,RGJ85} developing a small $\epsilon=4-d$ expansion for the study
  of finite-size effects in $d$-dimensional systems that are finite in all, or
  in all but one, of the $d$ fundamental directions and satisfy periodic
  boundary conditions. For background and references on finite-size effects,
  see Refs.~\onlinecite{Bar83}, \onlinecite{Pri90},
  \onlinecite{BDT00}, and \onlinecite{ZJ02}.}
\bibitem[{rem({\natexlab{b}})}]{rem:nabla}
\bibinfo{note}{The form~\eqref{eq:J} of the action $\mathcal{J}$ involving
  $\overleftarrow{\nabla}\cdot\overrightarrow{\nabla}$ is obtained directly if
  one keeps carefully track of the boundary contributions to $\delta
  \mathcal{H}= \int_{\mathfrak{V}}\{\delta\bm{\phi} \cdot[-\nabla^2 +\mathring{\tau}
  +(\mathring{u}/3!)\phi^2]\bm{\phi}\} +\int_{\mathfrak{B}}\delta\bm{\phi} \cdot(\mathring{c}
  -\partial_n) \bm{\phi}$ when going over from the Langevin equation to the
  Langrangian formulation. Rewriting the term
  $\int_{\mathfrak{V}}\tilde{\bm{\phi}}
  \overleftarrow{\nabla}\cdot\overrightarrow{\nabla}\bm{\phi}$ in terms of
  $\int_{\mathfrak{V}} \tilde{\bm{\phi}} (-\nabla^2) \bm{\phi}$ requires an
  integration by parts, which produces the boundary contribution
  $\int_{\mathfrak{B}}\tilde{\bm{\phi}}(-\partial_n) \bm{\phi}$. This
  contributes to the boundary terms of the classical equations of motion which
  yield the boundary conditions~\eqref{eq:bc}. The form of the action given in
  equation~(10) of Ref.~\onlinecite{GD06b} is consistent with our
  Eq.~\eqref{eq:J} provided contributions $\propto \delta(z)$ and $\propto
  \delta(z-L)$ are included in the functional derivative
  $\delta\mathcal{H}/\delta\bm{\phi}$ that yield the boundary terms
  $\int_{\mathfrak{B}_j} \tilde{\bm{\phi}}\cdot(\mathring{c}_j -\partial_n) \bm{\phi}$.
  For details, see Refs.~\onlinecite{DJ92} and \onlinecite{Die94b}.}
\bibitem[{rem({\natexlab{d}})}]{rem:Qd2}
\bibinfo{note}{The functions $Q_{d,2}$ are encountered also in the analysis of the quantum spherical model; see Ref.~\onlinecite{CPT98}.}
\bibitem[{rem({\natexlab{d}})}]{rem:DDG06}
\bibinfo{note}{Note that there is a misprint in the second part of equation (4.72) of Ref.~\onlinecite{DDG06}: the argument of the exponential function under the square root should be $\sqrt{\check{r}_\infty}$ rather than $\sqrt{\check{r}_\infty}/2$.}
\bibitem[{rem({\natexlab{d}})}]{rem:subllri}
\bibinfo{note}{As discussed in Ref.~\onlinecite{DDG06}, the linear
  scaling field associated with the subleading long-range interaction
  is coupled  beyond linear order to other linear scaling fields such
  as the one $\sim (u-u^*)$ associated with the usual Wegner-type
  corrections to scaling. This coupling implies a mixing of these
  irrelevant scaling fields whereby the nonlinear scaling field
  associated with the latter corrections to scaling acquires a
  dependence on the strength of the  subleading long-range
  interaction.}
\end{thebibliography}

\end{document}